\documentclass[twocolumn]{aastex631}
\usepackage[T1]{fontenc} 

\usepackage{mwe, tikz}

\usepackage{graphicx}
\usepackage{microtype}
\usepackage{url}
\usepackage{amsmath}
\usepackage{amssymb}
\usepackage{mathtools}
\usepackage{natbib}
\usepackage{multirow}
\usepackage{graphicx}
\usepackage{url}
\usepackage{xspace}

\usepackage[utf8x]{inputenc}

\newcommand{\ie}{{\textit{i.e.},~}}
\newcommand{\eg}{{\textit{e.g.},~}}

\newcommand{\figref}[1]{{\xspace}Fig.~\ref{#1}}

\newcommand{\secref}[1]{{\xspace}Sec.~\ref{#1}}
\renewcommand{\d}{{\mathrm{d}}}

\newcommand{\lmax}{\ell_\mathrm{max}}
\newcommand{\fsky}{f_\mathrm{sky}}
\renewcommand{\deg}{\mathrm{deg}}
\newcommand{\nside}{\textsc{nside}}
\newcommand{\healpix}{\textsc{Healpix}~}

\begin{document}
\raggedbottom\sloppy\sloppypar\frenchspacing

\vspace*{-1cm}
\title{
	Spurious correlations between galaxies and multi-epoch image stacks in the DESI Legacy Surveys 
} 

\author{Edgar Eggert}
\affil{Department of Physics, Imperial College London, \\Blackett Laboratory, Prince Consort Road, London SW7 2AZ, UK}

\author[0000-0002-3962-9274]{Boris Leistedt}
\affil{Department of Physics, Imperial College London, \\Blackett Laboratory, Prince Consort Road, London SW7 2AZ, UK}
\email{b.leistedt@imperial.ac.uk}


\begin{abstract}
A non-negligible source of systematic bias in cosmological analyses of galaxy surveys is the on-sky modulation caused by foregrounds and variable image characteristics such as observing conditions.
Standard mitigation techniques perform a regression between the observed galaxy density field and sky maps of the potential contaminants.
Such maps are ad-hoc, lossy summaries of the heterogeneous sets of co-added exposures that contribute to the survey.
We present a methodology to address this limitation, and extract the spurious correlations between the observed distribution of galaxies and arbitrary stacks of single-epoch exposures.
We study four types of galaxies (LRGs, ELGs, QSOs, LBGs) in the three regions of the DESI Legacy Surveys (North, South, DES), which results in twelve samples with varying levels and type of contamination.
We find that the new technique outperforms the traditional ones in all cases, and is able to remove higher levels of contamination.
This paves the way for new methods that extract more information from multi-epoch galaxy survey data and mitigate large-scale biases more effectively. 
\end{abstract}

\section{Introduction}

Over the last decades, the study of the statistical properties of galaxies has become a pillar of observational cosmology. 
Giant surveys of the night sky help us test increasingly sophisticated models of dark matter, dark energy, gravity, and galaxy formation and evolution.
One of the challenges of such surveys is to maintain systematic biases to sufficiently low levels.

One potential bias is the dependency of the data on observing conditions (\eg seeing), calibration (zero-points), and foregrounds (\eg stars, Galactic dust). 
This phenomenon is well-known and often comes into consideration when planning observations, as it is unavoidable that the data will not be exactly uniform \citep{Shafer_2015, Awan_2016, Almoubayyed_2020}.
This is sometimes referred to as the \textit{transfer function} of the survey, and the resulting spurious correlations in the galaxy numbers can mimic cosmological imprints and bias the downstream inference. 
Hence, they must be mitigated, as demonstrated in recent cosmological analyses by the Dark Energy Survey (DES, see \eg \citealt{Gatti_2021, Rodr_guez_Monroy_2022, Wagoner_2021}) and the Kilo Degree Survey (KiDS, see \eg \citealt{Johnston_2021, https://doi.org/10.48550/arxiv.2110.06947}).

One solution to this problem is to map out the potential contaminants on the sky, and ignore the corresponding modes when extracting the cosmological signal from galaxy number counts. 
This \textit{mode (de)projection} is effective but computationally expensive. 
A more tractable approach is to remove the (best-fit) correlations between the galaxy number counts and the contaminants at the leverl of maps of estimated power spectra.
While this is in principle a simple regression\footnote{This process must be done for each galaxy sample under consideration, since the transfer function would be different, as illustrated in this paper.}, deciding on the form of the model and its inputs is critical. Typically, the inputs include any property of the survey which could affect galaxy measurements: astrophysical foregrounds (\eg stellar density, dust extinction), observing conditions of the survey (\eg seeing, airmass, background noise), etc. 
The most common form of model is linear in the inputs, and predicts the galaxy number counts or the overdensity field.
More complex models (\eg based on neural networks) can provide significant improvements for some galaxy samples (see \eg \citealt{Rezaie_2020, chaussidon2021angular}).
Omitting inputs or adopting an overly simplistic model will lead to an incomplete subtraction of the spurious non-cosmological correlations. 
On the contrary, excessively complex models can lead to over-fitting and over-subtraction, because of random ``chance'' correlations between the inputs and the cosmological signal\footnote{Mode projection is not subject to this effect since it is equivalent to a Bayesian marginalization of the fit parameters.}.
However, this effect can be predicted analytically,  simulated, or mitigated by preventing over-fitting (as we do here).

Finally, a powerful avenue for modeling the transfer function is to employ synthetic source injections in images.
They play an important role in current surveys \citep{Everett_2022, Huang_2017}, and this role will grow as systematic biases must be modeled at higher accuracy.
However, they are very computationally demanding, so they likely complement the regression methodologies covered here.

This paper deals with a specific aspect of this problem: the multi-epoch nature of modern surveys. Indeed, to reach sufficient depth, surveys take multiple exposures covering the same sky area, to be co-added to reduce noise.
The exposures never align exactly and are not taken under the exact same observing conditions, which is further complicated by random artefacts such as contamination by cosmic rays, stars, or bad exposures generally.
As a result, any particular pointing on the sky is covered by a complicated stack of images, further increasing the complexity of the transfer function.
All aforementioned methodologies and previous works perform a compression (at each sky location) into a set of summary statistics (\eg the mean), resulting in maps of potential contaminants (abbreviated as \textit{contaminants maps} in the remainder of this paper).
Removing correlations between those maps and the galaxy number counts neglects the complexity of the underlying image stacks.
For example, co-added exposures are often diverse and include outliers, \ie single-epochs with extreme properties with respect to the rest of the other overlapping exposures, or the rest of the survey. To our knowledge, their effect on the uniformity of the transfer function and the accuracy of the aforementioned techniques is unknown.
This paper seeks to address this issue by employing a model architecture, deep sets, that is able to exploit heterogeneous image stacks. 

Deep sets \citep{zaheer2017deep, korshunova2018bruno, wagstaff2019limitations, murphy2019janossy, Soelch_2019, histogram_pooling,  kosiorek_2020, bronstein2021geometric} are a type of neural network capable to learn functions over un-ordered sets. 
They have found a range applications in physics and astronomy (\eg \citealt{https://doi.org/10.48550/arxiv.2007.04459, Komiske_2019}). 
In practice, we will modify the standard deep sets architecture in order to support the geometric setting we operate under, \ie variable number of un-aligned exposures. 
The metadata of the exposures (seeing, airmass, etc) form the basis of the data from the stacks we use to predict the galaxy number counts.

We analyse four types of galaxies (luminous red galaxies, emission line galaxies, quasars, and Lyman break galaxies) extracted from DESI Legacy Surveys (DLS) data, and illustrate the effect of the new technique on clustering measurements and cross-correlations with potential contaminants.
This allows us to follow several previous works that have analysed DLS galaxy samples, \eg \cite{Rezaie_2020, Kitanidis_2020, chaussidon2021angular, Zarrouk_2021}.
However we do not extract cosmological parameters from those measurements since they require more thorough tests (as well as realistic mocks) which must be different for each galaxy sample.
We also divide the sky into three areas, corresponding to the North, South, and Dark Energy Survey regions of the DLS data, which have significantly different properties. 
Overall, this gives 12 different population, each of with varying degrees of contamination, giving a wide range of examples to evaluate the performance of the cleaning methods.

With a methodology able to work with exposure stacks, one can circumvent the limitations of lossy contaminants maps. One may also be able to perform less stringent sky or data quality cuts.
This is particularly relevant for the arrival of data from the Legacy Survey of Space and Time (LSST) taken at the Vera C. Rubin Observatory, which will run over ten years and will have many more exposures than the DLS \citep{lsst_sciencereq}.
While the increased number of exposures can decrease the impact of outliers on image stacks, it cannot remove the spatial fluctuations altogether.
And in spite of optimized observing strategies being developed to minimize their effect on cosmological inference (\eg, \citealt{Awan_2016, Almoubayyed_2020}), at the level of precision of the LSST the techniques above will remain relevant.

This paper is structured as follows. 
In \secref{sec:data} we describe the data. 
\secref{sec:methods} covers the models.
We compare them in \secref{sec:results}, and conclude in \secref{sec:conclusion}.

Finally, note that we use \healpix to sub-divide the sky into pixels of resolution \nside=512, each containing 16 sub-pixels (resolution \nside=2048). 

\newpage
\section{Data}\label{sec:data}

This section describe the data set we use and the steps to prepare them for our comparison.

\subsection{The DESI Legacy Imaging Surveys (DLS)}

We use galaxy catalogs and image metadata from the Data Release 9 (DR9) of the DLS \citep{Dey_2019}. Those data are collected with multiple instruments.
In the North Galactic Cap, the Beijing-Arizona Sky Survey (BASS) covbered $5.100~\deg^2$ in the $g$- and $r$-bands using the Bok telescope located at Kitt Peak in Arizona \citep{Zou_2017}. This was complemented by $z$-band exposures from the The Mayall $z$-band Legacy Survey (MzLS) covering the same area as BASS with the Mayall telescope, which is also located at the Kitt Peak complex \citep{Zhou_2018, 10.1117/12.2231488}. These two telescopes make up the \textit{North} area of the DLS. 

A further part of the DR9 data comes from the Dark Energy Camera Legacy Survey (DECaLS), which covers the sky in all three bands $g$, $r$, and $z$ using the Dark Energy Camera (DECam) mounted on the Blanco Telescope at Cerro Telolo in Chile. The area covered by DECaLS covers almost $15.000~\deg^2$ of sky.
However, the part of the DECaLS catalogue that was used to generate the Dark Energy Survey (DES) features more exposures per pixel than the non-DES part of DECaLS. Hence, the DECaLS catalogue is further split into a {DES} part ($ 4600~\deg^2$) and a non-DES part ($9900~\deg^2$), with the latter referred to as \textit{South} from here onwards.
This separation is made based on a total number of exposures (in all bands) covering each $\nside=2048$ \healpix pixels, with DES having more than 60.

The DLS are complemented (as part of the DR9) with near-infrared fluxes extracted from the unWISE catalogues based on data from the WISE satellite \citep{meisner2017deep, wright2010wide}.

\subsection{Galaxy catalogs} 
We generate galaxy catalogues from the public DLS DR9 \textit{bricks} data. One data brick covers a pre-defined area of the sky and lists all objects detected in it, from the three ($g, r, z$) deep images made by co-added single-epoch images in this region. 
For each detected object, a host of metadata is available, including location in right ascension (R.A.) and declination (Dec.), fluxes, and Milky Way transmission, among other statistics. 
On average, every brick contains around 9.000 objects, and there are more than 350.000 unique bricks in DR9. 
We process every brick with the \textsc{desitarget} code\footnote{\url{https://github.com/desihub/desitarget}}.
The catalog making step already includes a sophisticated masking of bad pixels or objects near bright stars or extended galaxies, which has been refined over the years \citep{Kitanidis_2020, Rezaie_2020, chaussidon2021angular, Zarrouk_2021}.

We consider four samples of galaxies:  \textbf{Luminous red galaxies (LRGs)},  \textbf{Emission line galaxies (ELGs)}, \textbf{Quasars (QSOs)}, and \textbf{\textit{g}-dropout Lyman Break Galaxies (GLBGs)}. 
The colour cut for the first three galaxy types can be found in the \textsc{desitarget} pipeline as well as in \cite{Kitanidis_2020, Rezaie_2020, chaussidon2021angular}, for example. 
We only needed to define cuts for the GLBGs (analogous to that of quasars): signal-to-noise ratios smaller than 3 in the $r$ band (\ie a non-detection) and greater than 4 in the $r$, $z$, and $W1$ bands, and greater than $3$ or $2$ for the $W2$ band (in the South or North, respectively). Those are only detection cuts and do not include the color cuts typically added to remove contaminants from Lyman Break selection \citep{Hildebrandt_2009, Harikane_2017, Harikane_2022}.
The same quality mask as the ELGs (`{notinELG}') is applied.

We generate number counts sky maps at \healpix resolution $\nside=512$. 
Higher resolutions are more challenging due to the larger number of pixels, the smaller number of galaxies in each pixel and the associated Poisson noise. 

\subsection{Image properties (exposure-level contaminants)} 

The first type of potential contaminants are the properties of the exposures that were processed as part of the DLS data (making the co-added images covering the bricks) and from which galaxy catalogs are constructed.
We extract this information from the seven million annotated exposures of the DLS DR9. 
The metadata of the exposures form the basis of the data from the exposure stacks we use to predict the galaxy number counts.

Geometrically speaking, exposures cover the sky in a complicated manner, and as a result any sky pointing is described by a stack of exposures (different in the three $grz$ bands).
In order to resolve and store this information, we identify which exposures cover (the centers of)\footnote{Focusing on the centers of \healpix pixels rather than resolving the full geometry of the problem is an approximation sufficient for this work.} \healpix $\nside=2048$ pixels, and store the metadata in a database-like format, since the variable length of the stacks makes it impossible to store this in rectangular arrays without resorting to zero-padding.

\begin{figure}
    \centering
    \includegraphics[width=\columnwidth]{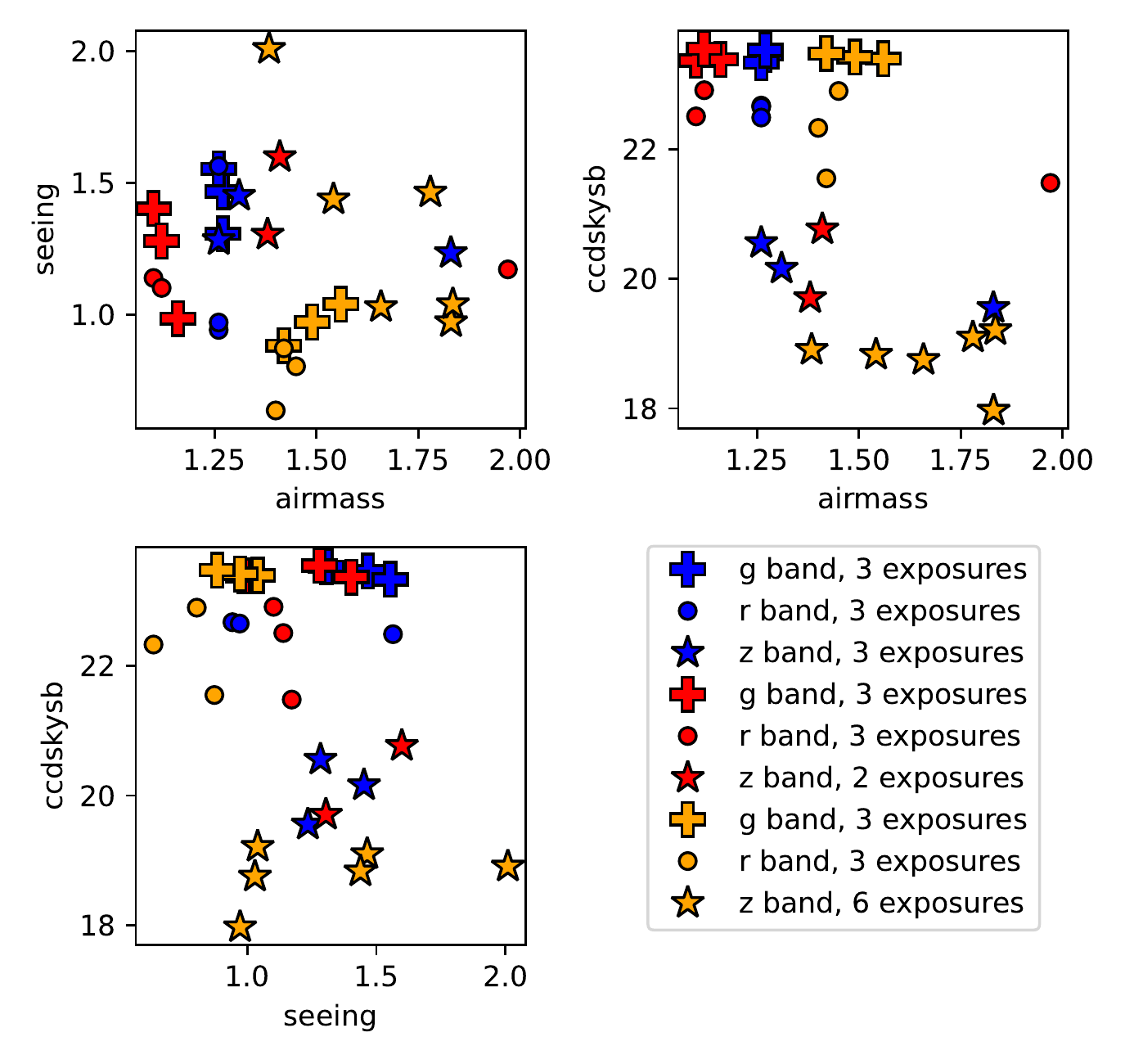}
    \caption{Illustration of the diversity of the properties of single-epoch exposures as well as their variable length. A subset of properties are shown for the exposures covering three $\nside=2048$ sub-pixels (ids 21975319, 27803830, and 791280). We develop a method to predict the contamination of galaxy catalogs from those permutation-invariant, variable-length stacks of exposure properties, rather than from the summary statistics (e.g., mean over exposures) employed by conventional methods.}
    \label{fig:ccd_stacks}
\end{figure}

For each exposure, we extract the following properties: the seeing, airmass, sky surface brightness, and sky counts (airmass is averaged over bands, as in \citealt{Kitanidis_2020}).
They are readily available in the publicly released DLS tables, so no image processing is required. 
They are also likely to be effective summaries of the images.
However, other choices would be possible (including a direct processing of the images).
The diversity of image stacks is illustrated in \figref{fig:ccd_stacks}.
We further compress them into maps by averaging the values in the stack of each pixel. 
This is the information employed by conventional systematics mitigation techniques.
We also calculate the standard deviation, minimum, and maximum, of each vector, which we will use for additional null-tests.

\subsection{External maps of contaminants (sky-level systematics)} 

Some contaminations does not originate from the observations (exposure) themselves, but rather from Galactic foregrounds, and therefore depend on the sky location of a given pixel. We call these sky-level contaminants and store them as \healpix maps at $\nside=512$.

We include a map of galactic dust extinction, extracted using the galactic dust map SFD98 \citep{schlegel1998maps, dustmaps}. We also include maps that trace the density of stars on the sky, which can be mistaken as galaxies and enter our catalogs as spurious objects, or affect the density or properties of detected galaxies (\eg the tails of bright stars). 
Two maps (`GAIA12' and `GAIA') are based on the Gaia data \citep{gaiadr2_2018}: the density of point sources in the magnitude ranges $\textsc{phot}\_G\_\textsc{mean}\_\textsc{mag} < 12$  and $12 < \textsc{phot}\_G\_\textsc{mean}\_\textsc{mag} < 17$.
We also follow the procedure of \cite{chaussidon2021angular} to make a map of the Sagitarius stream from the catalogs of \cite{Antoja_2020}. 
Finally, we make an additional stellar map generated by counting point-source objects from the bricks, following \cite{Kitanidis_2020}. 
Outliers were cleaned by removing all objects more than 3 standard deviations away from the stellar locus, calculated as the median color as a function of magnitude.
Additionally, we include a map of neutral hydrogen column density (`HINH'), from \cite{2016HI4PI}. 

More discussion on the origin of each contaminant (both exposure-level  and sky-level), and how they affect catalogs, can be found in \cite{Kitanidis_2020, chaussidon2021angular}.

Table~\ref{contaminants_list} is a summary of all the potential contaminants included in this work.

\begin{table}[htp!]
    \centering
    \begin{tabular}{|c|c|c|}
        \hline
        Contaminant & Level  &  X \\ 
        \hline
        Stellar & Pixel & multiband \\
        EBV & Pixel & multiband \\
        HINH & Pixel & multiband \\
        GAIA & Pixel & multiband \\
        GAIA12 & Pixel & multiband \\
        Sagittarius & Pixel & multiband \\
        Airmass &Exposure  & multiband \\
        Seeing &Exposure  & g,r,z \\
        Sky Surface Brightness &Exposure  & g,r,z \\
        CCD Skycounts &Exposure  & g,r,z \\
        \hline
    \end{tabular}
    \caption{Systematic contaminants and the level at which they impact galaxy density.}
    \label{contaminants_list}
\end{table}

\subsection{Sky cuts} 

We create DR9 coverage masks (at $\nside=512$ and $2048$) from the random catalogs available\footnote{\url{https://www.legacysurvey.org/dr9/files/}}. Pixels with insufficient coverage (<95\%) are removed. 
Furthermore, all objects situated in the Large Magellanic Cloud between R.A. $
\in$ [52° , 100°] and Dec $
\in$ [-70°, -50°], are cut, since this area is usually too heavily contaminated by stars to draw reliable inferences on the galaxy distribution. 

We then split the remaining sky into the three different areas (North, South, and DES) using the number exposures (DES having more than 60 exposures at $\nside=2048$) and cuts on the R.A. and Dec. 

We subsequently apply further sky cuts based on the maps of the previous sections in order to reject the most extreme values. They can be caused by processing issues, or simply by the most extreme image qualities. This also has the benefit to confine the values taken by each map to a reasonable range, suitable for machine learning algorithms once mapped to $[0, 1]$.  
We cut pixels with values outside of the median plus and minus ten times the median absolute deviation.

\begin{figure}
    \centering
    \includegraphics[trim=0 2.7cm 0 0, clip, width=\columnwidth]{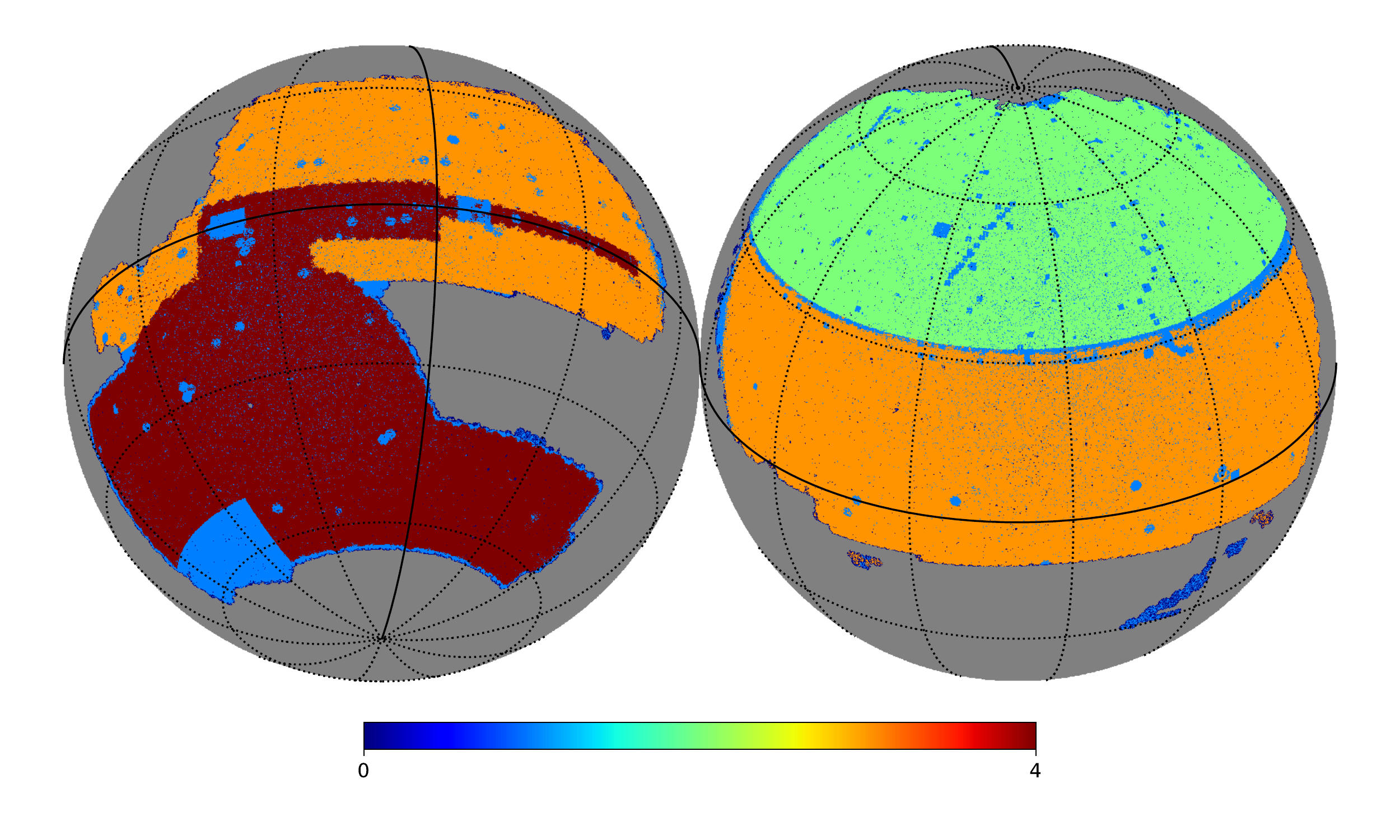}
    \caption{Sky coverage of the three regions used in this work (distinguished by color). Additional sky cuts (indicated in light blue) are performed based on the coverage of the survey as well as statistics of the various potential contaminants (e.g., pixels with extreme values were cut).}
    \label{fig:mask}
\end{figure}

The final three regions are shown in \figref{fig:mask}, and the average numbers of objects are in Table~\ref{tab:nbs}. 
The variations between regions are an indication of the differences in selection cuts, average depth, or spatial fluctuations. 

\begin{table}[htp!]
    \centering
    \begin{tabular}{|c|c|c|c|}
    \hline
         & North   & South   & DES    \\
         \hline 
        Sky fraction [\%]  & 0.115 & 0.211 & 0.108\\
        \hline 
        $\mathrm{N}_\mathrm{LRG}/\mathrm{arcmin}^2$  & 610.0 & 610.0 & 610.0 \\
        $\mathrm{N}_\mathrm{ELG}/\mathrm{arcmin}^2$  & 2363.9 & 2440.1 & 2363.9 \\
        $\mathrm{N}_\mathrm{QSO}/\mathrm{arcmin}^2$  & 305.0 & 305.0 & 228.8 \\
        $\mathrm{N}_\mathrm{GLBG}/\mathrm{arcmin}^2$  & 2211.4 & 1525.1 & 2135.1 \\
        \hline
    \end{tabular}
    \caption{Average (over each masked area) number of objects per surface area. Differences are due to variations in depth or observing conditions, or in the color cuts between the North and South regions.}
    \label{tab:nbs}
\end{table}

\section{Methods}\label{sec:methods}


\subsection{Map-based linear corrections model} 

If the contaminants are summarized in \healpix pixels (\ie compressed into maps), which is the standard approach, then modeling the spurious fluctuations amounts to running a simple regression to predict galaxy number counts from the values of potential contaminants (in pixels).
A linear model is the simplest model one can employ, and it has been widely developed in the literature (\eg  \citealt{Ross_2011, Kitanidis_2020, Elvin_Poole_2018, Wagoner_2021}) and successfully applied to recent surveys.
An extensive review is provided in \cite{Weaverdyck_2021}, where connections between different modeling choices and training strategies are also provided.
Given that this model is not backed by physics, there is freedom in the choice of scaling of the inputs and outputs, and also in the loss function to optimize.
We perform the fit using Python’s \textsc{Scikit-learn} package \citep{scikit-learn}. We found that ridge and lasso regressions did not provide any improvements over ordinary least squares linear regression, which we thus adopt.
Min-Max Scaling is applied to scale all inputs into $[0, 1]$ to facilitate model training.  
We do not scale the output. Even though it may be slightly more challenging to fit the galaxy number counts, not scaling it allows us to more easily compare metrics between the models and sky areas.

\subsection{Map-based non-linear corrections model} 

Opting for a non-linear function of the potential contaminants can provide significant improvements over the linear model and is routinely employed. 
Common choices include random forests and neural networks \citep{Rezaie_2020, Zarrouk_2021, chaussidon2021angular}. 
We adopt the latter.
It will serve as a baseline to compare the performance of the deep sets.
Min-Max Scaling is also applied to input contaminants, and no scaling to the output predictions. 

\subsection{Exposure-based deep-sets non-linear corrections model}

The previous methods rely crucially on compressing the contaminants of all the images of a pixel image stack into a one-dimensional feature vector (here the mean seeing, airmass, CCD counts and background noise in all three bands, with airmass averaged over bands).
We now explore whether one can model the contamination directly from the properties of the CCDs that cover a given pixel.
This would remove the lossy step of compressing the exposure properties into maps, and potentially extract more information from the full stacks of exposures. 
There are two complications arising: the sets of exposure vectors vary in size, and they have no inherent ordering.

The problem of learning a function on unordered, variable-sized sets is not new. Examples from other fields include 3D-point clouds \citep{qi2017pointnet} or population-level statistics \citep{zaheer2017deep}. Earlier attempts to learn functions on sets include the Neural Statistician of \citep{edwards2016towards}.  \cite{zaheer2017deep} was the first proof of concept for an architecture specifically designed to satisfy the criteria outlined above. Since then, the theory and applications of permutation-invariant layers have blossomed \citep{ravanbakhsh2017deep, korshunova2018bruno, wagstaff2019limitations, murphy2019janossy, Soelch_2019, histogram_pooling, kosiorek_2020, bronstein2021geometric}, but deep sets remain the simplest approach.  The core idea behind a deep sets architecture is to include layers performing permutation-invariant operations on the inputs, \ie yielding the same result regardless of the sets ordering. The simplest and most widely-used choices for such functions include aggregating (sum) or averaging (mean) over the input elements within a set, or taking the minimum or maximum value.

Technically, a deep sets architecture consists of three distinct blocks. First of all, there is a block that applies transformations to all individual elements in the input set. In the present context, this block will be referred to as a \textit{feature extractor} and is simply a fully connected feed forward neural network. Afterwards, the learned representations of every element in the input set are passed into the permutation-invariant layer, a so called \textit{aggregator}. This layer will learn the weights to optimally aggregate the function representations that were transformed in the \textit{feature extractor}. The \textit{aggregator} compresses the inputs set to a single dimensional feature vector of fixed size. This feature vector is now passed to the third block, where it is reduced to a final continuous output to predict the number of galaxies in this pixel. This last block is another fully connected feed forward neural network, however, it has a single output neuron and linear activation in the final layer. The final block will be referred to as \textit{MLP}. 

The 16 features of the previous sections are now replaced with 5 exposure-level features (seeing, sky counts, airmass and surface brightness, as well as a categorical integer variable for the band) and the 6 other sky-level potential contaminant maps which do not vary between exposures (extinction, stellar maps, etc). 
As such, these features are only passed to the MLP block of the network after the exposure-variant systematics are passed through the feature extractor and the aggregator. This is a first difference with the classic deep sets architecture.
Furthermore, the task of learning the contamination in galaxy number counts presents additional challenges that prevents the use of the conventional deep sets architectures of \cite{lee2019set, zaheer2017deep, Soelch_2019, qi2017pointnet}. 

\paragraph{Variable set sizes}

First of all, the chosen architecture needs to account for variable sized inputs. Few implementations in the literature adapted their networks to process variable sized inputs, and for those that did, the variability of input set sizes was handled with zero-padding to a fixed input set size. However, the feature extraction network includes a bias term, meaning that the padded zero-vectors are problematically transformed to non-zero vectors. This would impact the outcome of the aggregation block that follows the feature extractor. Consequently, we need to include an additional \textit{masking} operation that tells the aggregator which elements of the input set to aggregate and which ones to ignore for the aggregation. This mask is generated for every input set independently and passed to the aggregator block during a forward pass trough the network.

While zero-padding followed by masking the zero-vectors usually solves the problem of variable set sizes, a further problem arises. The generated datasets feature very large variations in the number of exposures that are associated to a given pixel. That is, the set sizes that are fed into the network differ substantially across different pixels. To ensure that a given pixel meets a minimum quality threshold, only pixels that were covered by a minimum of two exposures per band are kept. Since there are three bands ($g$, $r$, $z$), some pixels on the sky are only covered by six exposures. However, other pixels are covered much deeper, especially in the DES area, with tens of exposures per band per pixel. Using large maximum set sizes substantially increases training time since the feature extraction layer is applied to every exposure vector, regardless of whether it represents an actual exposure or was added during zero-padding. 

\paragraph{Input Dimensionality}

 To overcome the problem of prohibitively large zero-padded vectors, we 
 employ a sub-pixelisation. When dividing each pixel into 16 $\nside=2048$ sub-pixels, each of is covered by fewer CCDs, reducing the maximum necessary set size that needs to be supported by the architecture. 
 Furthermore, fewer zero-vectors would be processed in the feature extractor, thus speeding up model training. However, we still want the output galaxy number counts to be calculated at resolution $\nside=512$. As a result, the model includes tensors with another input dimension (sub-pixels), and a final layer averaging them into pixels, diverging further from the standard deep sets architecture. The maximum set size is set such that less than 0.1\% of all subpixels in the footprint of this area are covered by more exposures. For the 1 in 1000 subpixels that featured more subpixels, a random sub-set of exposures are drawn. 
 The maximum set sizes are 30, 25, and 40, for the North, South, and DES regions, respectively.\\

\begin{figure*}
    \centering
    \includegraphics[width=\textwidth]{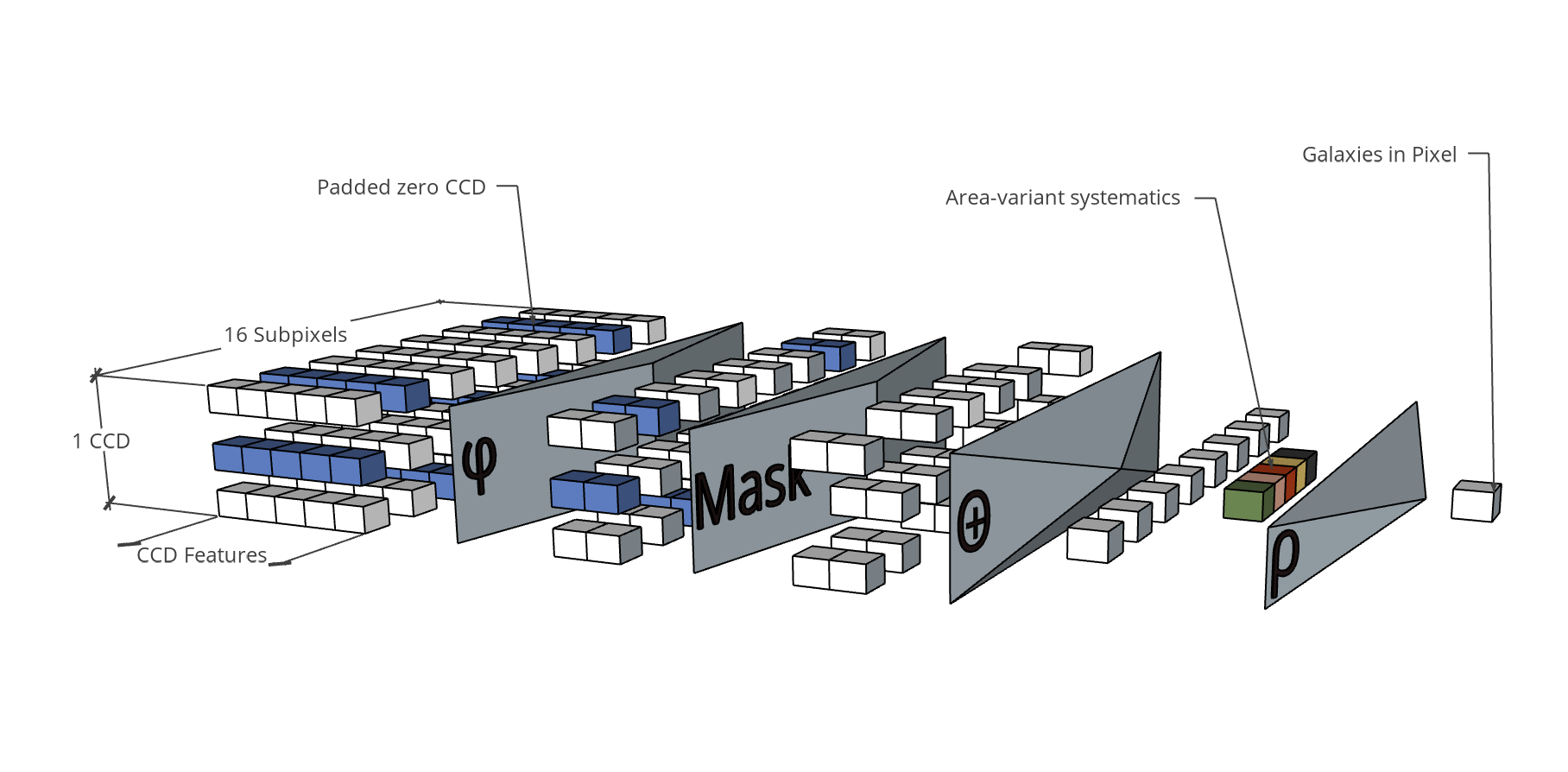}
    \caption{Schematic of the deep sets-inspired architecture employed in this work to support arbitrary image stacks. CCD features are the summary properties of each CCD (see Table 7, ten in total). Depth corresponds to the 16 $\nside=2048$ sub-pixels (some of which removed by the Mask if they don't fall in the valid region) contributing to each $\nside=512$ pixel, which are to be aggregated by a permutation-invariant operator (\eg the sum). The height corresponds to one 1 CCD, which may not cover all the sub-pixels. Sky-level systematics (extinction, stellar counts, etc, in Table 7) are appended, before a final prediction is calculated for a $\nside=512$ pixel. Phi and rho are parametrized with neural networks. }
    \label{fig:deep_set_arch}
\end{figure*}

A summary schematic of our deep sets architecture is shown in \figref{fig:deep_set_arch}. 
The values for the input CCDs for the deep sets were scaled using a Robust Scaler utilising the interquartile range, rather than the minimum and maximum. As previously, the output galaxy number counts per pixel were not scaled.

\subsection{Hyperparameter optimization} 

Hyperparameters can drastically affect the ability of a model to generalize to unseen data. 
K-fold cross-validation has previously been used on the non-linear model (\eg \citealt{Rezaie_2020}). While useful in scenarios with limited or highly imbalanced data, a major drawback of k-fold cross-validation is that all models now have to be trained $k$ times. We found that k-fold validation was not needed for the data considered here; random assignment of pixels to training and validation folds would still generate samples that are representative of one another.

The fully connected feed forward neural network and the deep sets were built in the \textsc{PyTorch} Python framework \citep{pytorch}. The software package \textsc{optuna} \citep{akiba2019optuna} is used to randomly draw values of the hyperparameters in their respective ranges of interest, run the optimization, and automatically stop poor runs (``pruning") and start new ones in other regions of the prior. 
This allows us to explore more effectively the high-dimensional space of hyper-parameters.

We now enumerate the various hyperparameters relevant here, and what ranges we explore. We first consider the parameters relevant to the neural network model.

We use a Poisson \textbf{loss function} \citep{Rezaie_2020}, which we optimize using the \textsc{Adam} \textbf{optimizer} \citep{kingma2017adam}. The \textbf{learning rates} for our models are sampled uniformly in $[10^{-5}, 10^{-2}]$. Additionally, we employ a \textbf{weight decay regularisation} to improve robustness. This is achieved by adding a function of the weights in the model to the loss function, which enforces weight sparsity and prevents the weights from becoming large. This is controlled by a hyperparameter $\lambda$ passed to the optimiser \citep{kingma2017adam}, sampled uniformly in $[0, 0.3]$. 
The multilayer perceptrons trained here are allowed to have between 1 and 6 layers. The hidden layers can take on a minimum of 8 neurons per layer (approximately half the feature space) all the way up to 256 neurons per layer, to also allow for wider networks. Those integer parameters are sampled uniformly.
We consider the following values for the \textbf{batch size}: 16, 32, 128 or 256. 
To improve robustness, feed-forward neural networks models can be trained by randomly dropping a given percent of neurons during training \citep{hinton2012improving}. The \textbf{dropout percentage} parameter was randomly sampled uniformly in $[0, 0.5]$, meaning that in each drop-out layer anywhere between zero to half the neurons are randomly deactivated during training. \\

For the deep sets, there are additional hyperparameters. First of all, a deep set consists of two fully connected feed-forward neural networks that need to be configured, the feature extractor in the beginning, and the MLP network at the end of the model. As such, parameters such as number of layers, neurons per layer and percent dropout have to be drawn and ultimately set for each network independently, using the same ranges as above. Furthermore, the deep set architecture can use different permutation-invariant functions in the aggregator block of the network. The choices here were limited to aggregation by using the maximum value of a given set, aggregation to the mean, or summing the values passed from the feature extractor. Lastly, the potential depth of a deep sets architecture can lead to the network getting stuck local minima during optimization. As such, several different initializations are available to the networks parameters. The choices here are Xavier-Glorot (XG), Kaiming He (KH), following a uniform or a normal distribution.

\begin{table}[ht]
\caption{Comparison of the results of the fits from the various models (with the best model in each row highlighted in bold). The metric is the coefficient of determination $R^2$ calculated over the full area (thus including the training, validation, and testing data). For each model, a hyperparameter search was performed. } 
\begin{center}
\begin{tabular}{|c|c|c|c|c|}
    \hline
    Area & Galaxy Type & Linear & Neural Net & DeepSet \\
    \hline
    \multirow{5}{*}{North} & LRG & 0.01 & 0.008 & {\bf 0.013} \\
    & ELG & 0.08 & 0.139 & {\bf 0.144} \\
    & QSO & 0.104 & 0.112 & {\bf 0.127} \\
    & GLBG & 0.109 & 0.21 & {\bf 0.261} \\
    \hline
    \multirow{5}{*}{South} & LRG & 0.006 & 0.006 & 0.006 \\
    & ELG & 0.061 & 0.108 & {\bf 0.129} \\
    & QSO & 0.091 & 0.095 & {\bf 0.1} \\
    & GLBG & 0.046 & 0.167 & {\bf 0.189}\\
    \hline
    \multirow{5}{*}{DES} & LRG & 0.011 & 0.02 & {\bf 0.026} \\
    & ELG & 0.027 & {\bf 0.048} & {\bf 0.048} \\
    & QSO & {\bf 0.035} & 0.033 & 0.03 \\
    & GLBG & 0.075 & 0.105 & {\bf 0.134} \\
    \hline
\end{tabular}
\end{center}
\label{tab:results}
\end{table}

\subsection{Maps and power spectra}\label{sec:maps_and_power_spectra}

To evaluate the performance of the cleaning methods, we will look at maps and angular power spectra.
When needed, number count maps can be converted to over-density maps by dividing by the average number density (and subtracting one). This must be done in each region separately.
Those overdensity maps are what we can calculate the angular spectrum of, for null tests or comparison with cosmological models (which we do not do here).
Angular power spectra are calculated using \textsc{NaMaster} \citep{Alonso_2019} applied to the \nside=512 maps, with spherical harmonic modes up to $\lmax=1024$, again in each region separately (even though the sky maps plotted in this paper show them together).
Individual multipoles $\ell$ are binned into bands of size $11$. All power spectra and covariances are calculated for this band size, as implemented in \textsc{NaMaster}.
When covariances on the measured (auto- or cross-) power spectra are needed, we use a Gaussian approximation  \citep{Alonso_2019}.
However, the latter requires a model for the underlying true power spectrum. 
We simply feed the measured power spectra (sum of measured spherical harmonic coefficients, before deconvolving the effect of the mask) divided by the fraction of sky covered. This is a rough estimate of the underlying power spectrum, which delivers positive and smoother power spectra than those resulting from full deconvolution of the mask. 
It is sufficient for the purpose of obtaining covariances which, in our case, are only used to 1) divide power spectra by their typical uncertainties to illustrate fractional chances, 2) calculate chi-squared statistics when performing null tests with cross-power spectra.
In particular, the same covariance is fed to the chi-squared of the three methods, so that they can be compared, and the accuracy of the covariance only affects the absolute value of the chi-squared, which is less critical in our case since we are concerned with the relative values before and after cleaning.

\begin{figure}[!htb]
    \centering
    \includegraphics[width=0.95\columnwidth]{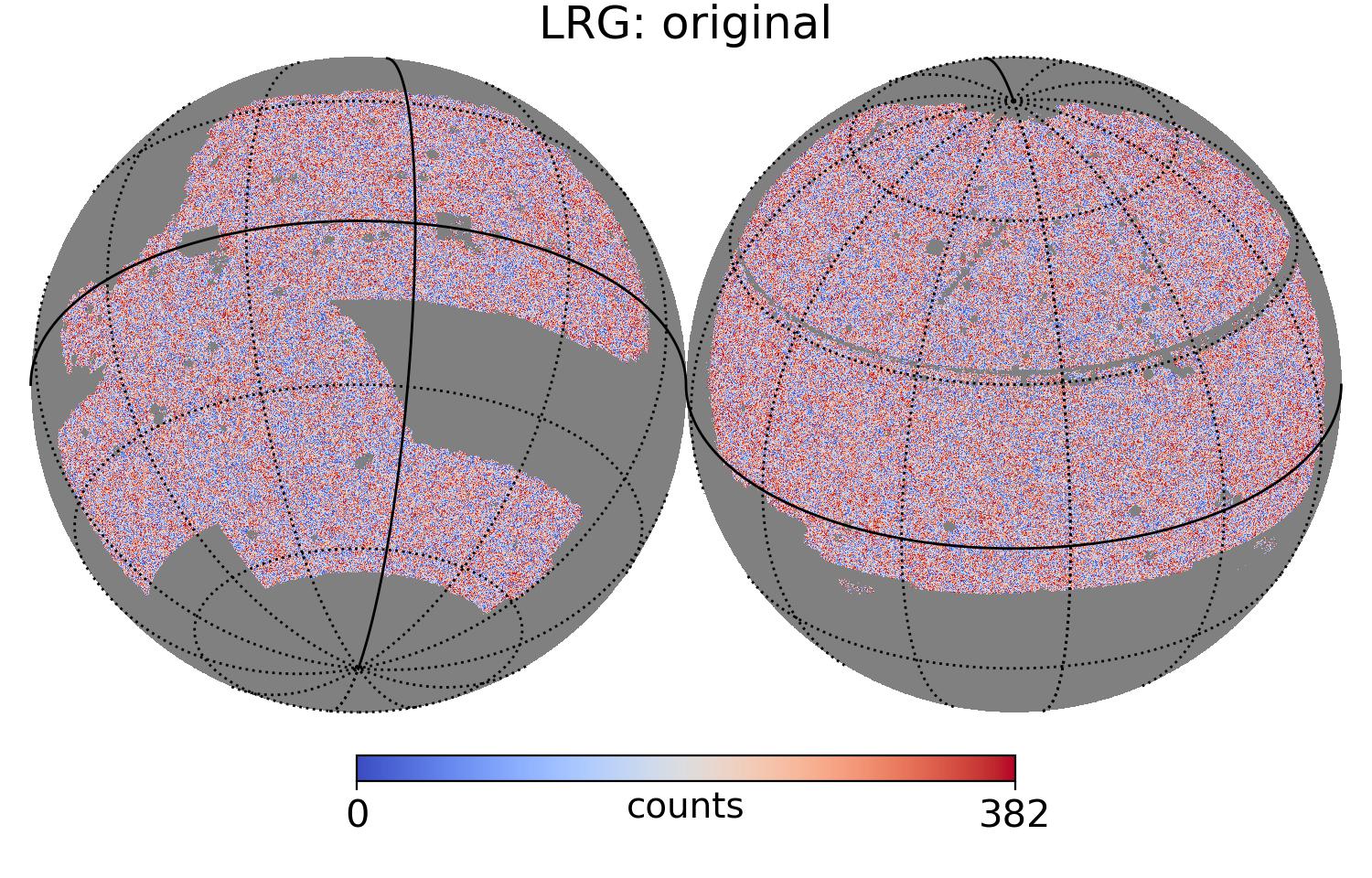}
    \includegraphics[width=0.95\columnwidth]{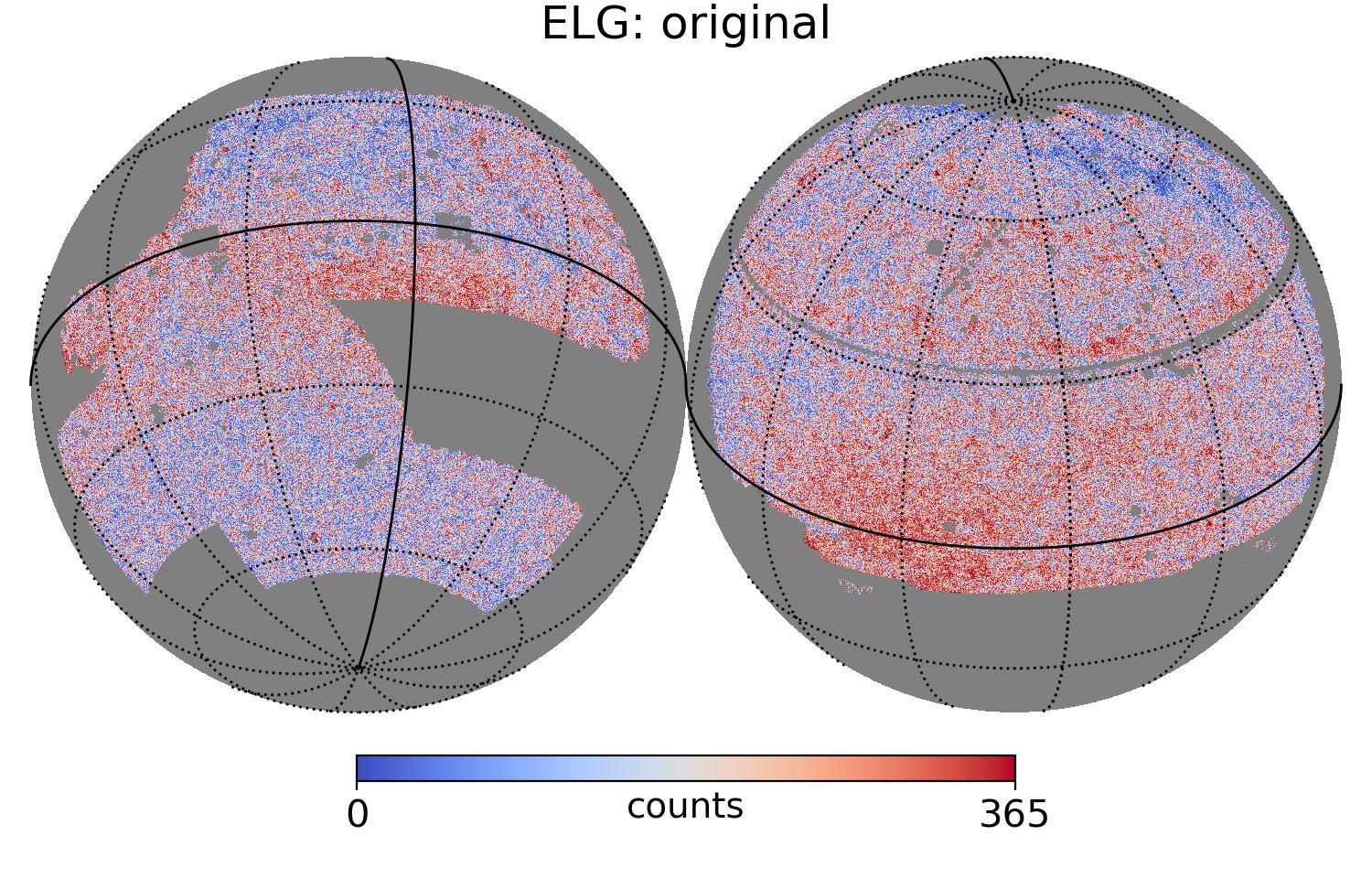}
    \includegraphics[width=0.95\columnwidth]{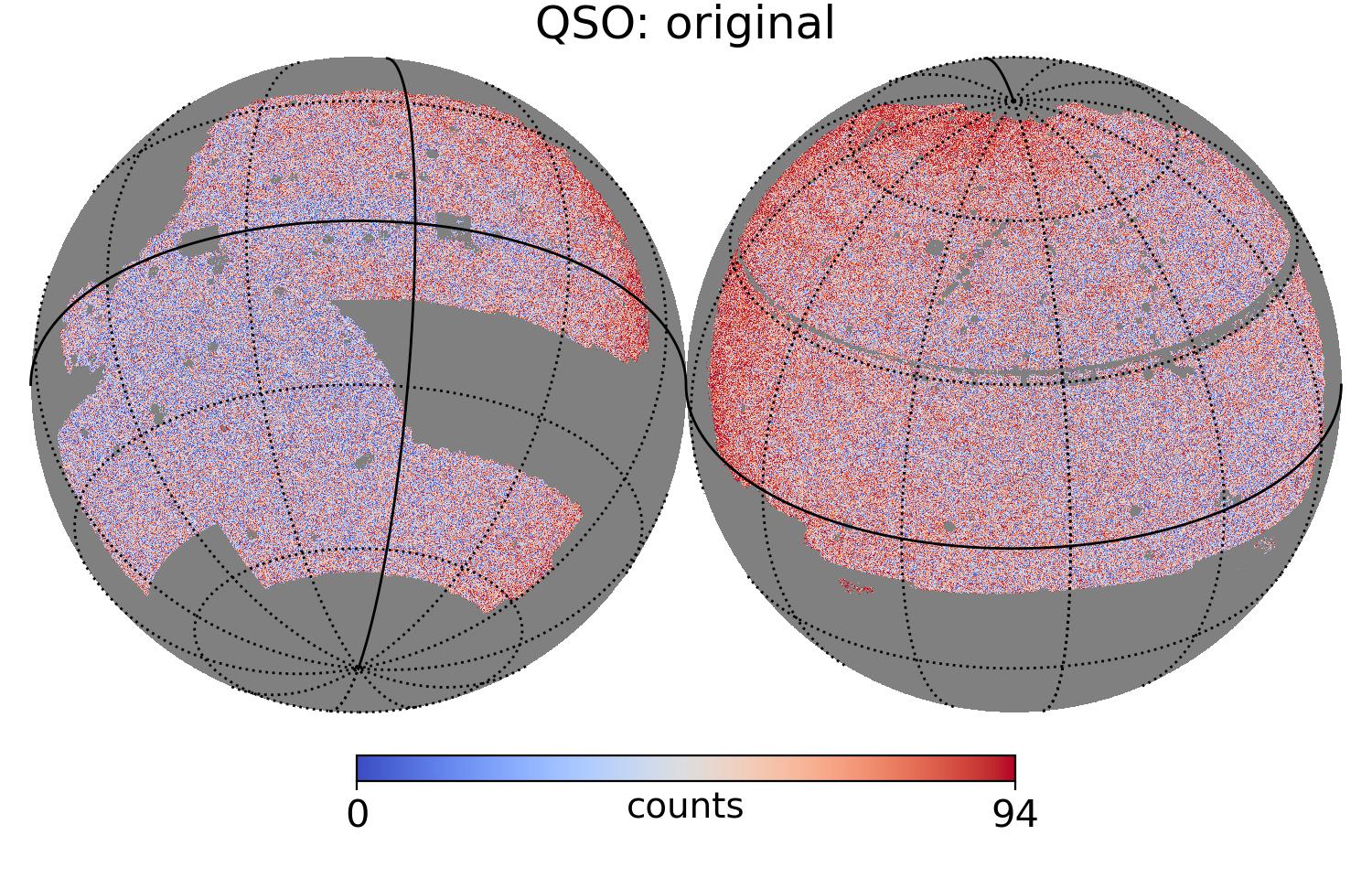}
    \includegraphics[width=0.95\columnwidth]{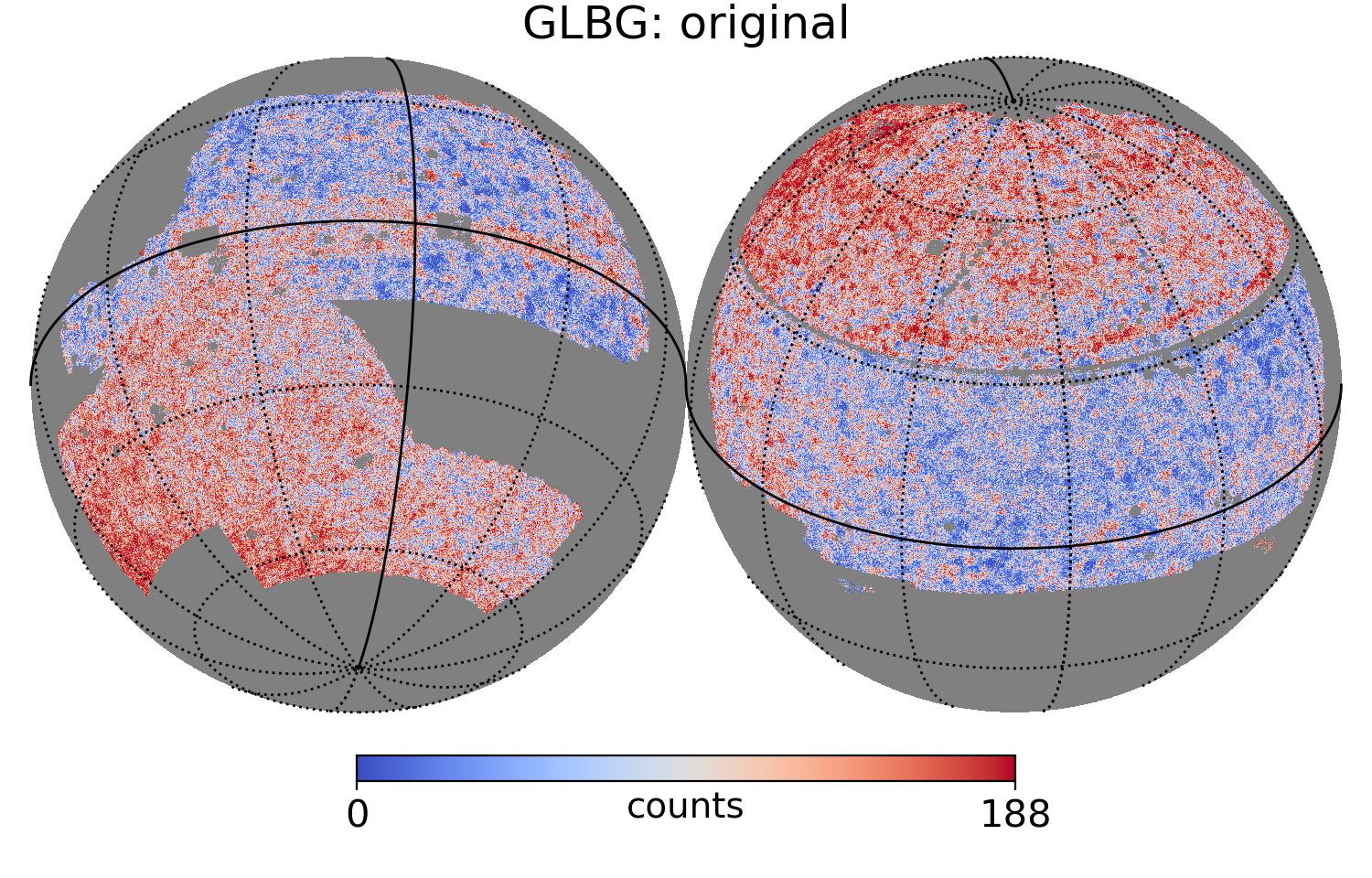}
    \caption{Original, uncorrected galaxy catalogs (number counts in \healpix pixels $\nside=512$) constructed from the DESI Legacy Surveys data DR9. Suspicious features on a variety of scales are present and most likely non-cosmological.}
    \label{fig:maps_orig}
\end{figure}

\begin{figure}[!htb]
    \centering
    \includegraphics[width=0.95\columnwidth]{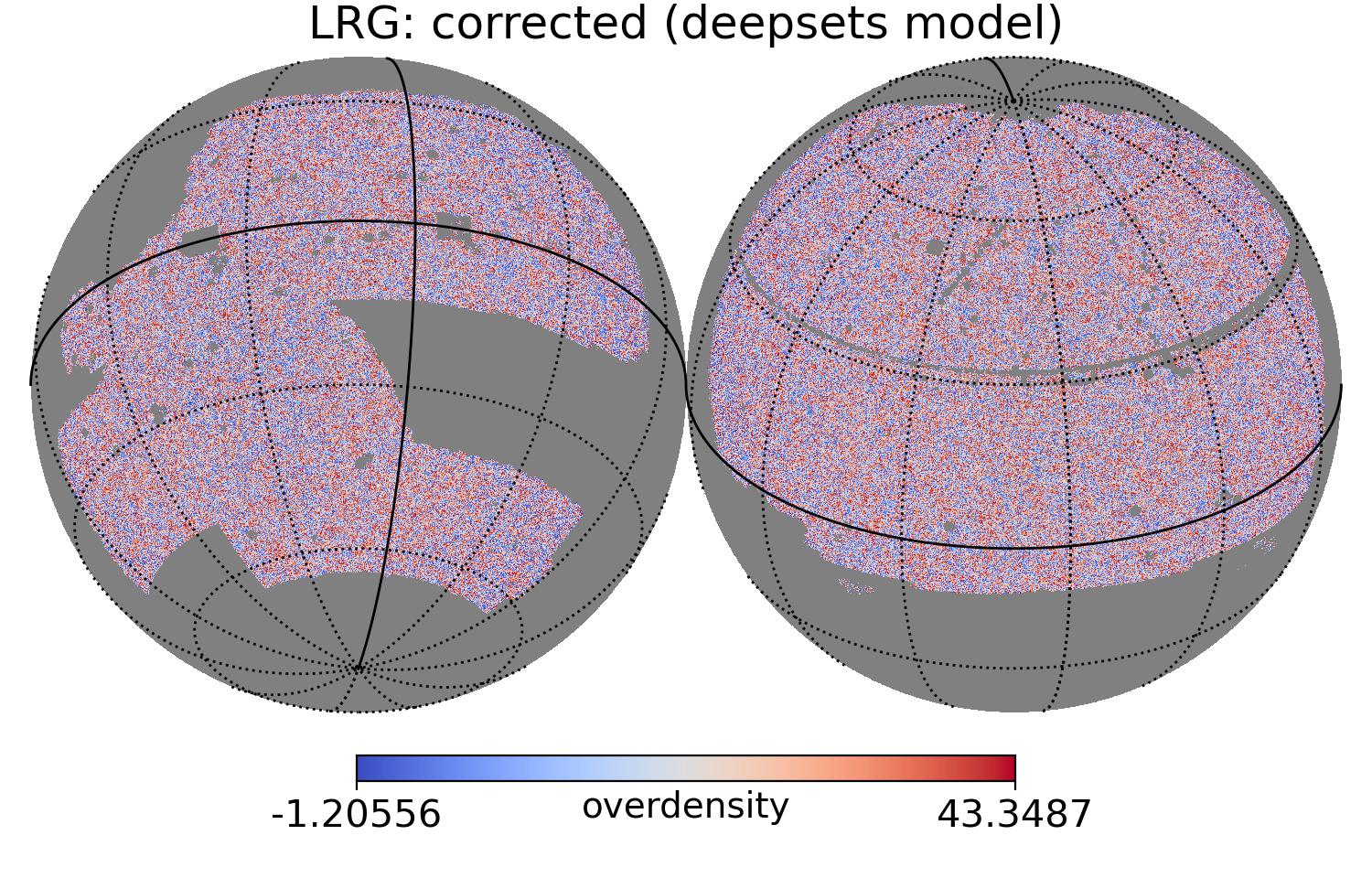}
    \includegraphics[width=0.95\columnwidth]{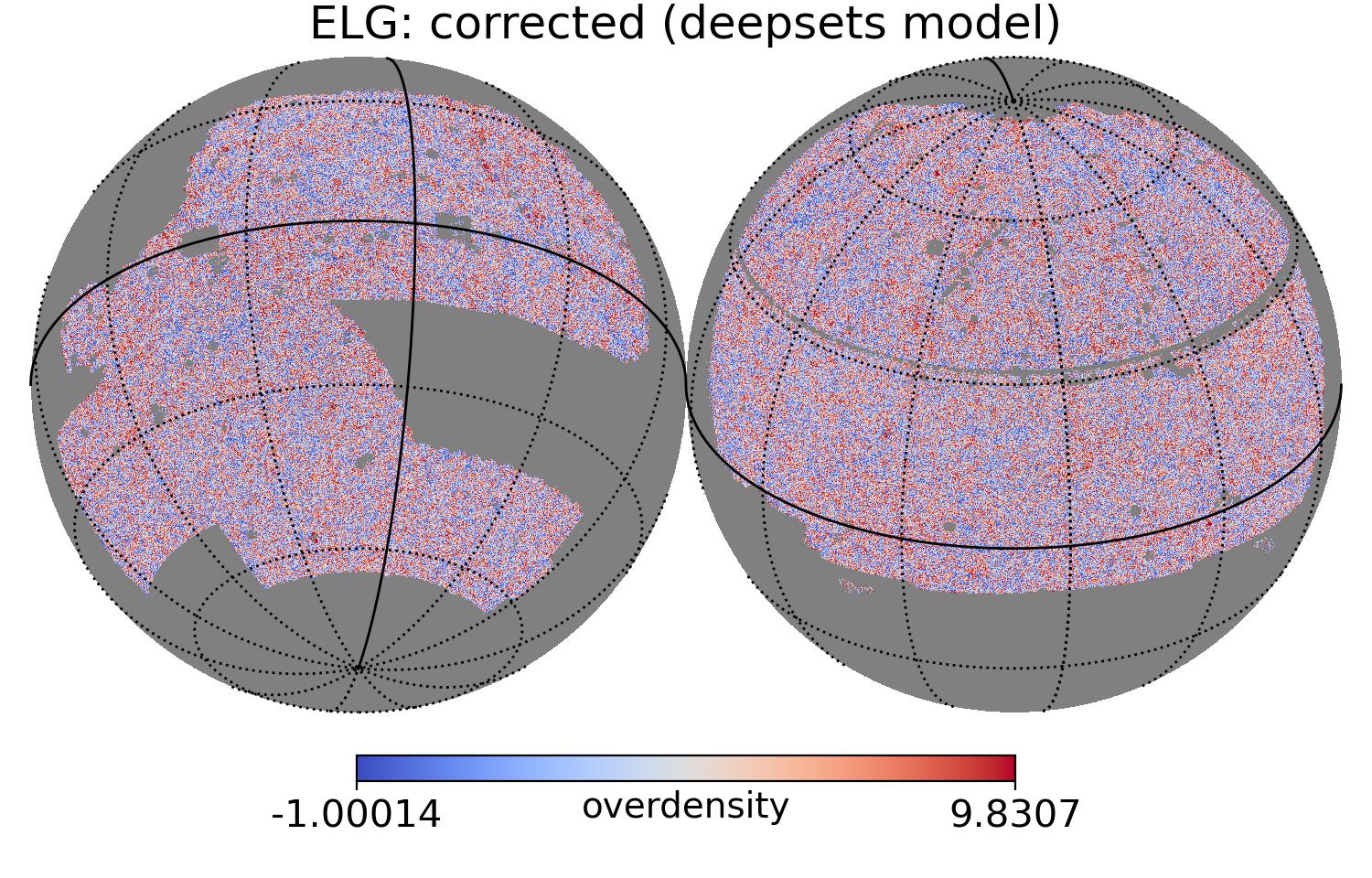}
    \includegraphics[width=0.95\columnwidth]{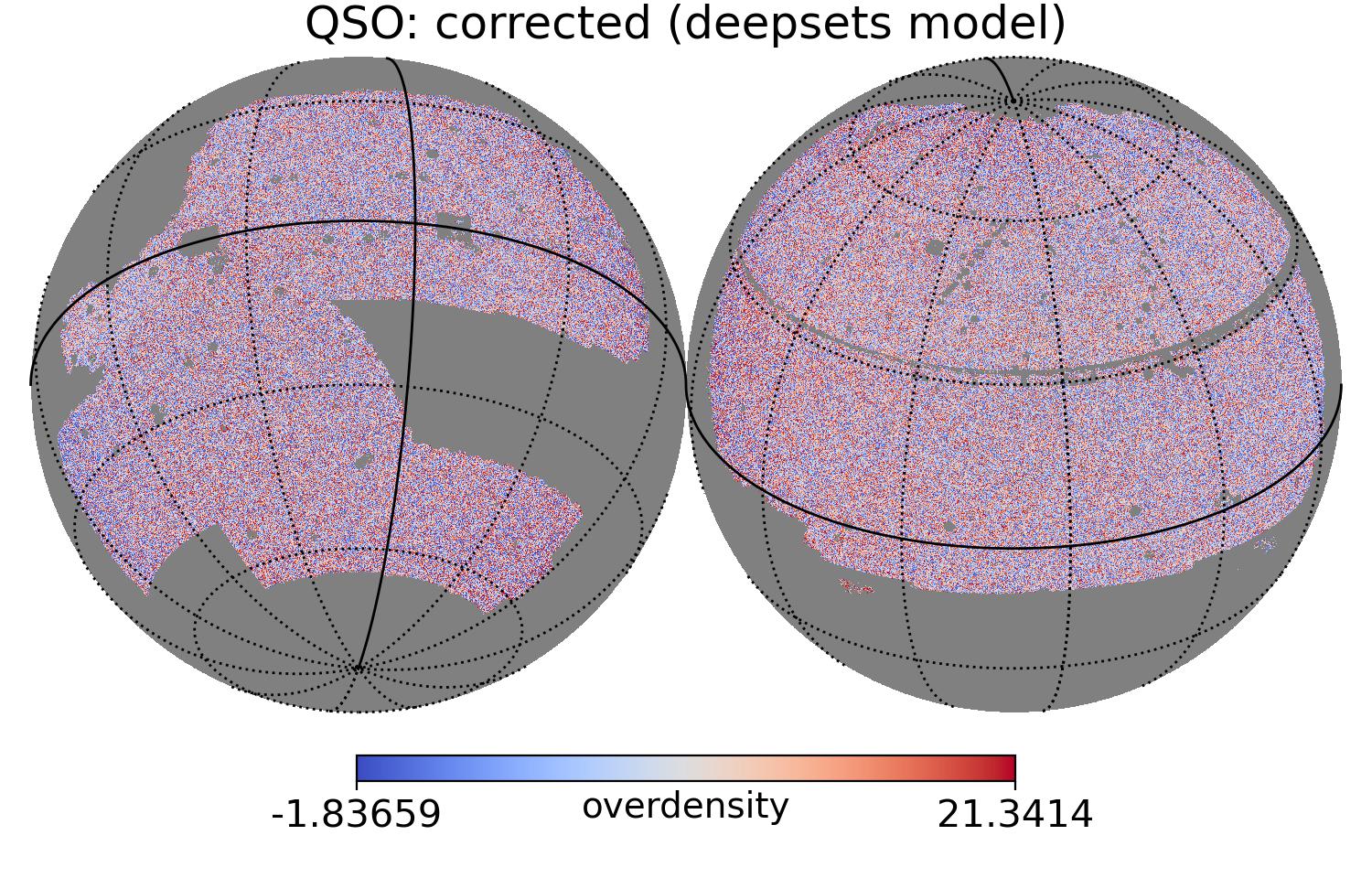}
    \includegraphics[width=0.95\columnwidth]{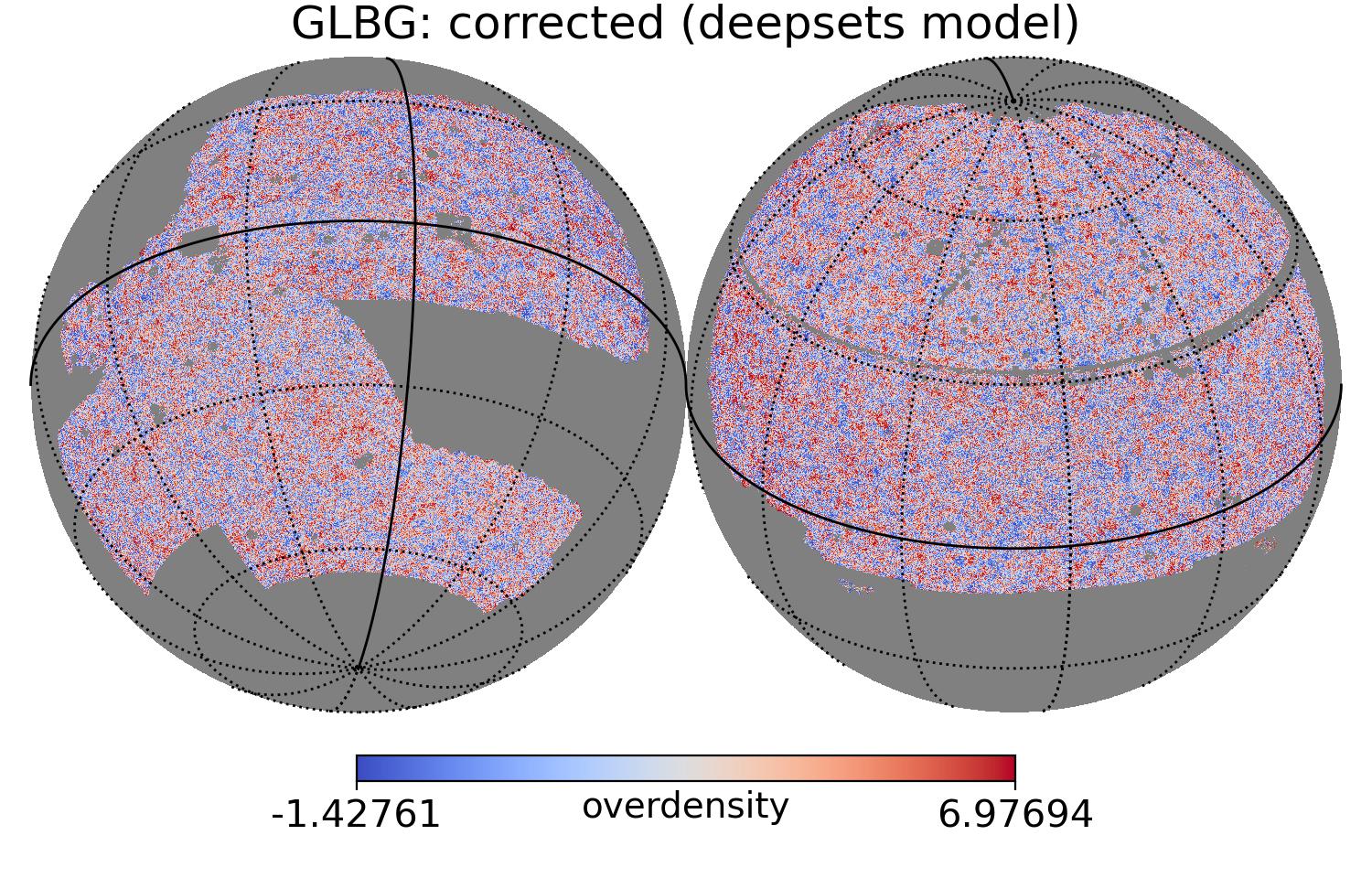}
    \caption{Number counts maps obtained after correcting for the correlations with potential contaminants using the new deep sets-based technique. Residual fluctuations are nearly isotropic and uniform across the sky.}
    \label{fig:maps_deep_corr}
\end{figure}

\begin{figure}[!htb]
    \centering
    \includegraphics[width=0.95\columnwidth]{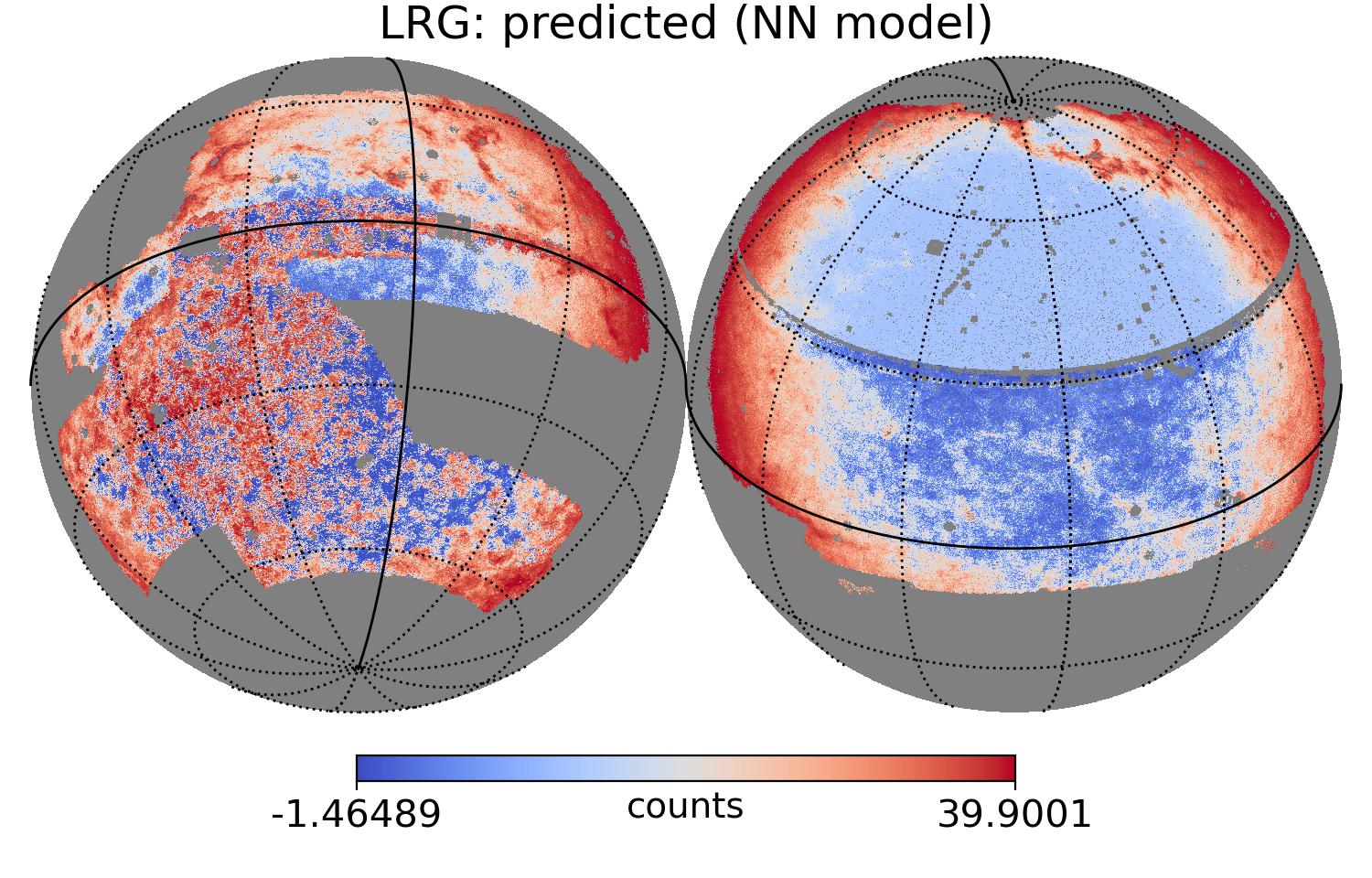}
    \includegraphics[width=0.95\columnwidth]{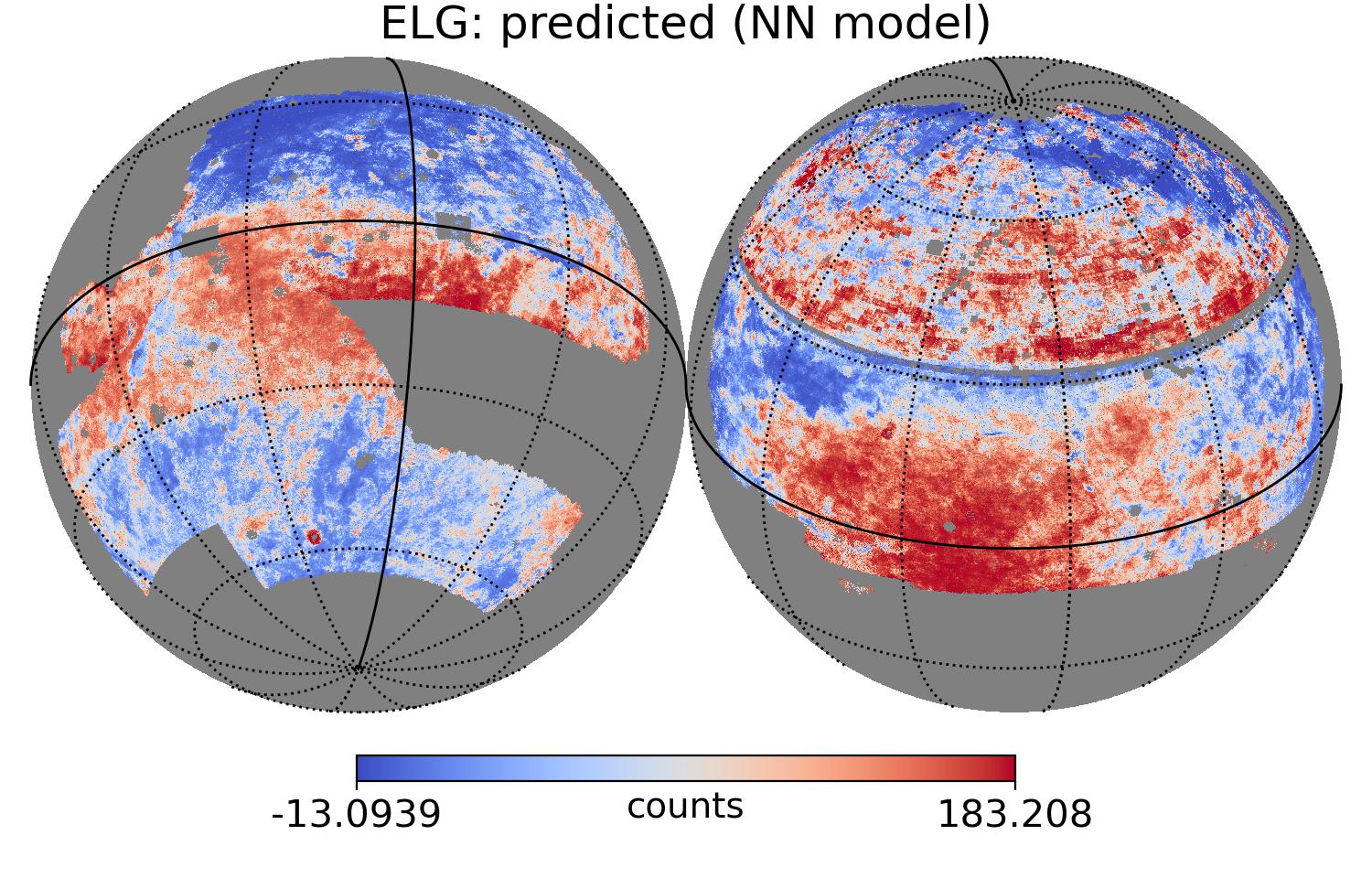}
    \includegraphics[width=0.95\columnwidth]{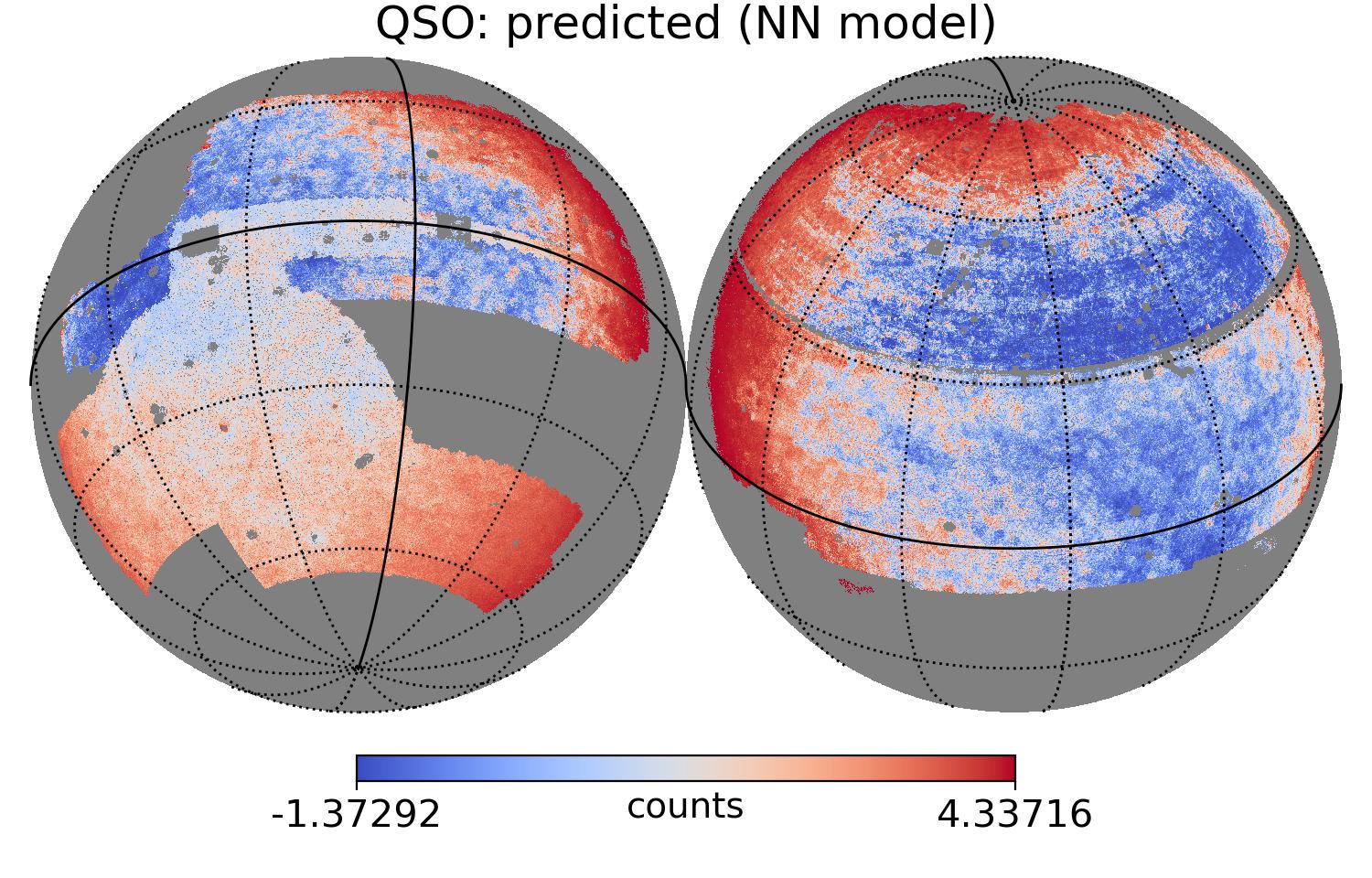}
    \includegraphics[width=0.95\columnwidth]{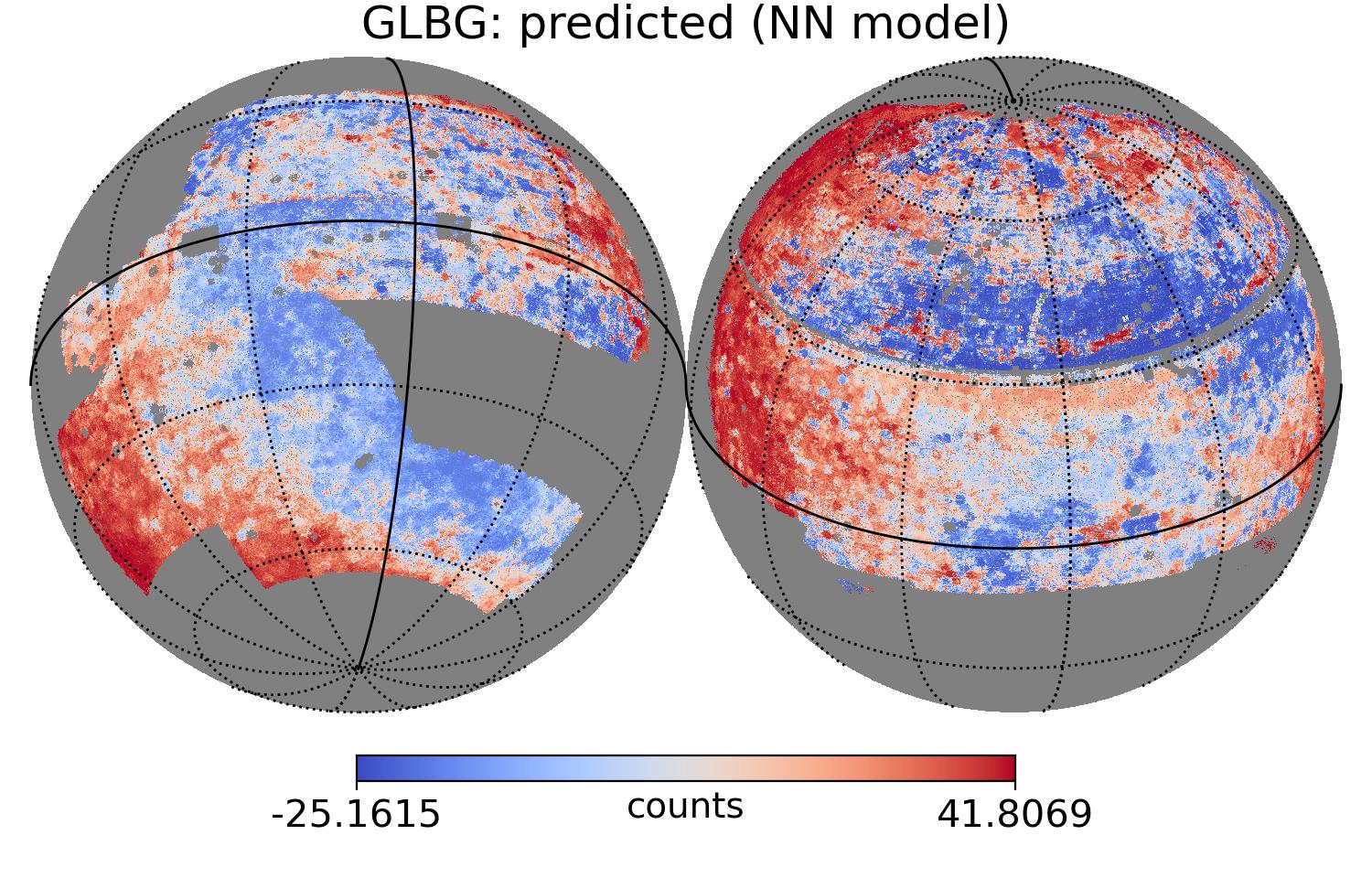}
    \caption{Correlations found by the neural network model between the original number counts maps of \figref{fig:maps_orig} and the maps of potential contaminants (which only include the mean seeing, airmass, ccd noise, over exposure stacks).}
    \label{fig:maps_nn} 
\end{figure}

\begin{figure}[!htb]
    \centering
    \includegraphics[width=0.95\columnwidth]{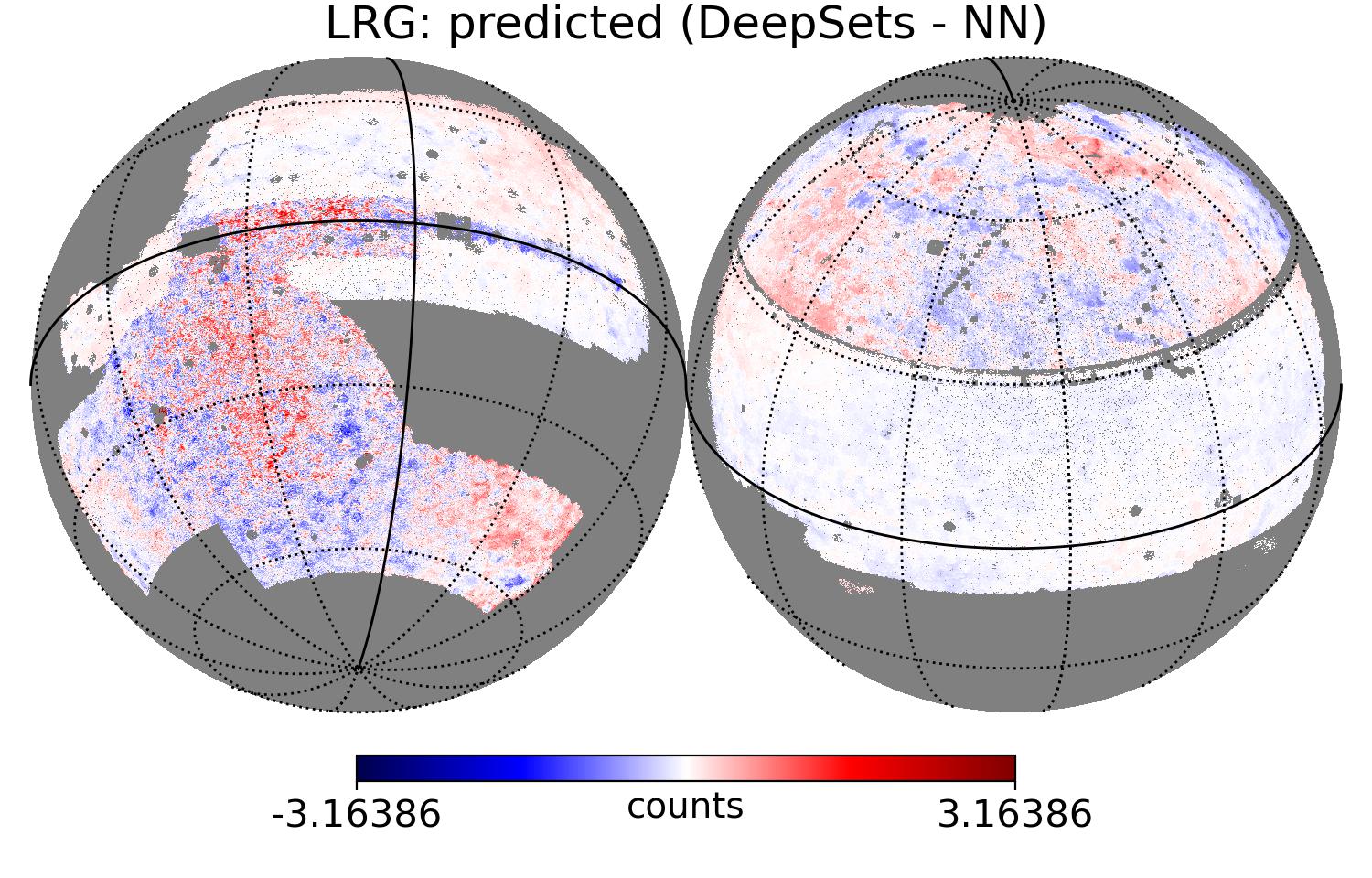}
    \includegraphics[width=0.95\columnwidth]{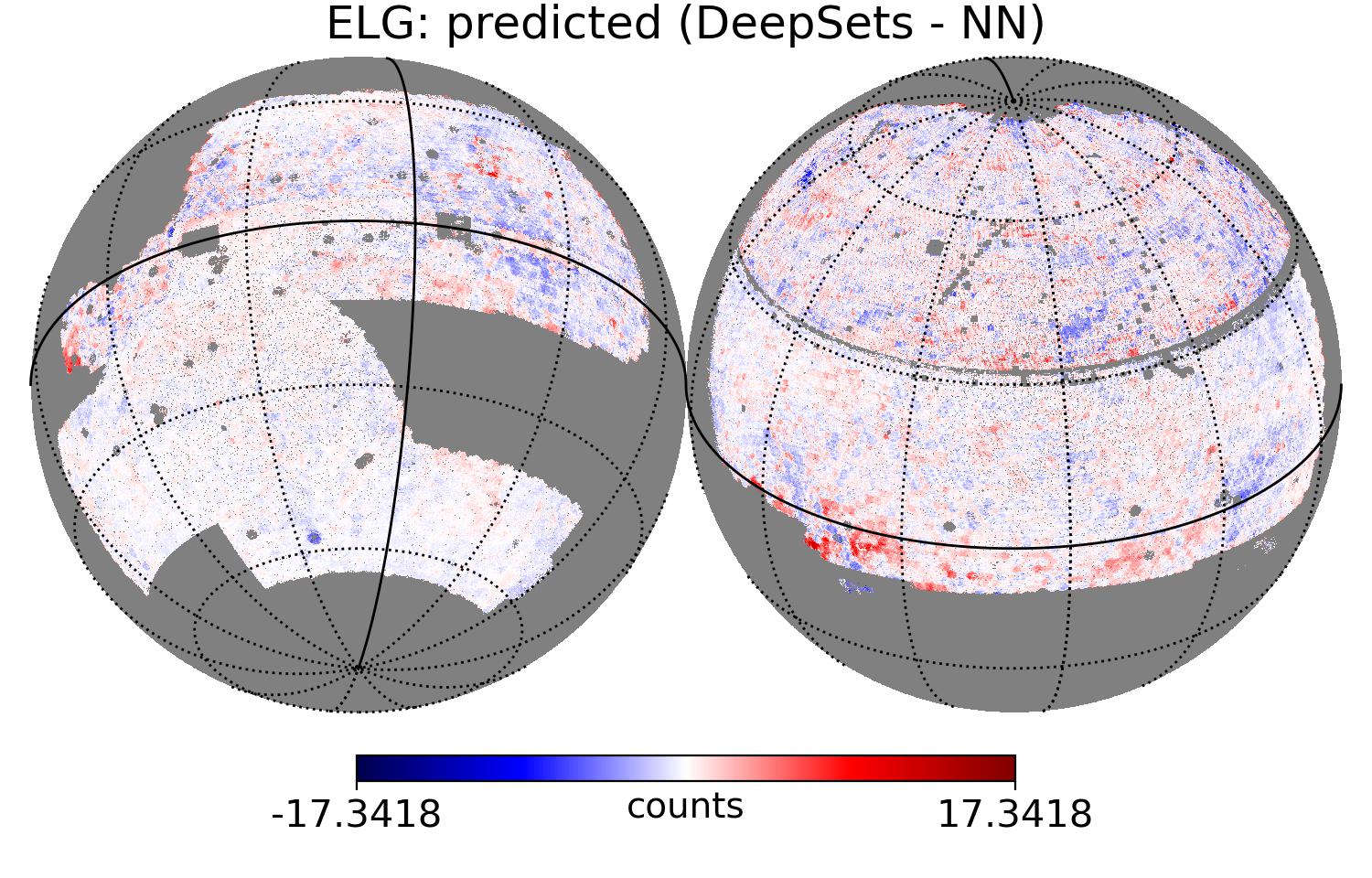}
    \includegraphics[width=0.95\columnwidth]{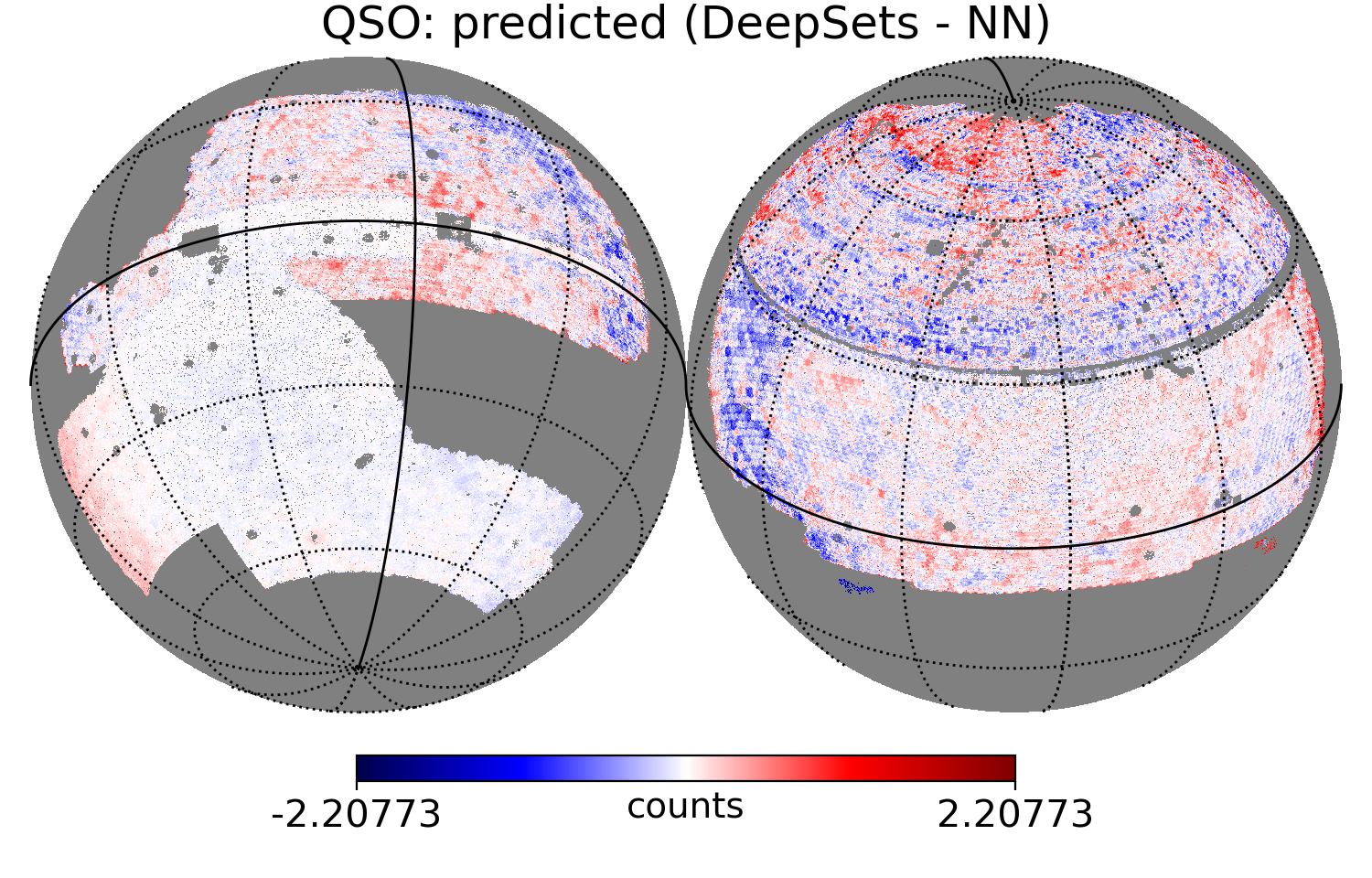}
    \includegraphics[width=0.95\columnwidth]{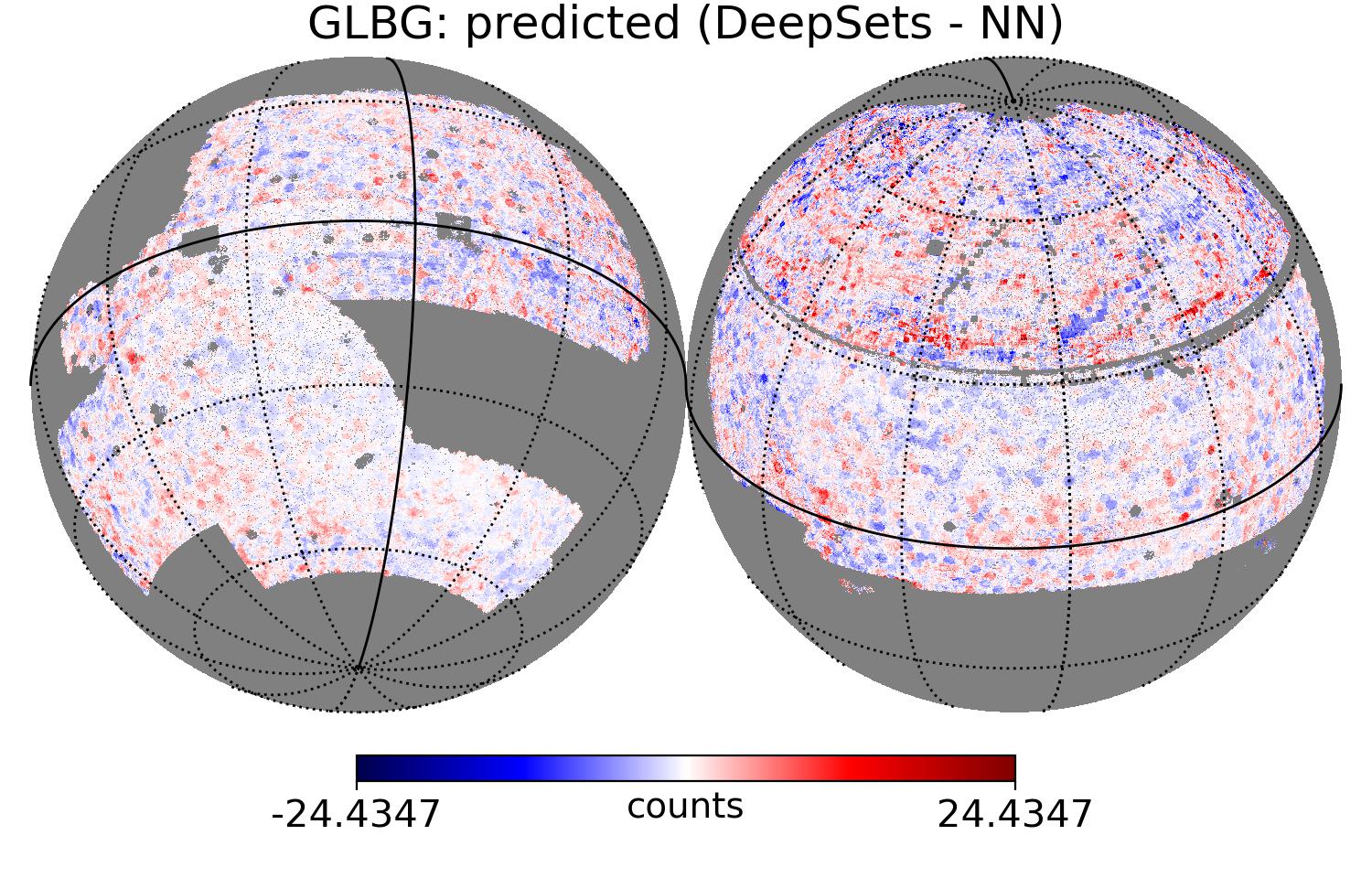}
    \caption{Additional spurious correlations found (on top of those shown in \figref{fig:maps_nn}) in the number count maps by the new deep sets model from the full exposure stacks.}
    \label{fig:maps_deep}
\end{figure}

\section{Results}\label{sec:results}

\subsection{Results of each fit} 

We randomly split the data in each region into training (60\%), validation (20\%), and testing (20\%) sets of pixels (thus not necessarily adjacent).
We train each model on the first set, and use the second for stopping when the model performance no longer improves.
For the non-linear and deep sets models, this is coupled with the exploration of hyperparameters performed with \textsc{optuna}.
We adopt the best models, as measured by the coefficient of determination $R^2$ calculated on the test data set.
Finally, we apply those models to the full data set (re-assembling the training, validation, and testing sets in each region), which we will use below.
The corresponding $R^2$ are shown in Table~\ref{tab:results}.
We do not discuss the hyperparameters in detail here\footnote{They can be found at \url{https://github.com/elleggert/astrostatistics/blob/master/models/deep_set/final_run.py}.} because they are not directly interpretable, so their values (in absolute terms, or relative to the other regions or models) do not provide additional insight.

We see in Table~\ref{tab:results} that except in cases with very low levels of contamination (\ie LRGs), the deep sets model outperforms the linear and neural network models.
This is not unexpected given that it has many more degrees of freedom and thus should be able to extract any contamination in the data more effectively.

\begin{figure*}[!ht]
    \centering
    \includegraphics[trim=1mm 4mm 1mm 3mm, clip, width=\textwidth]{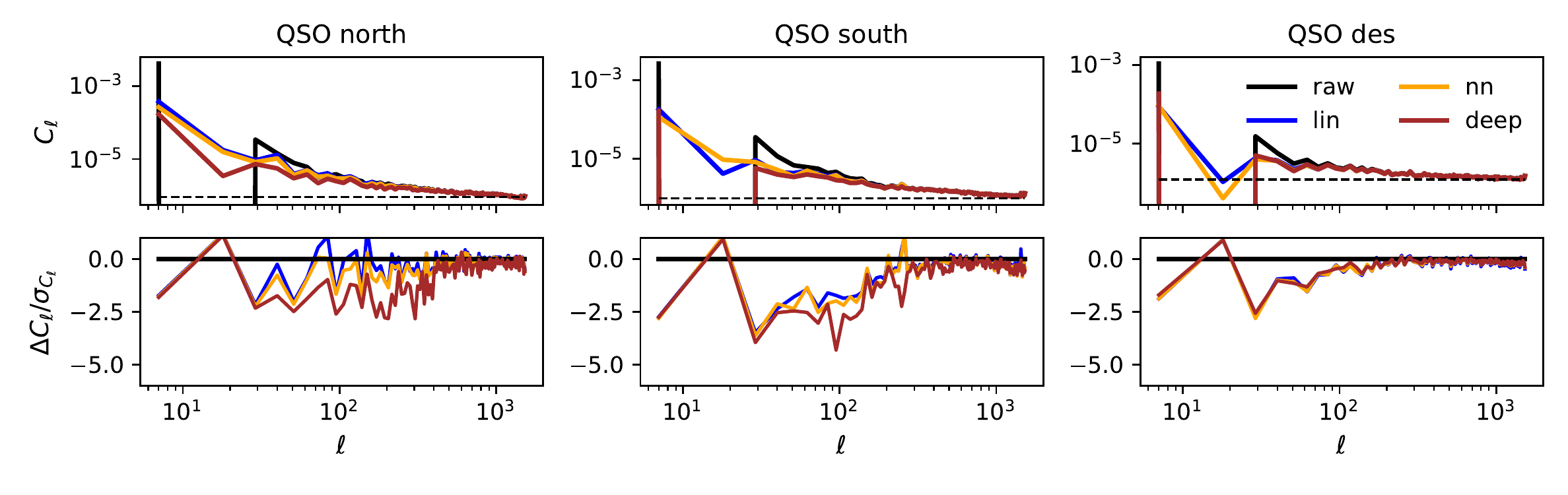}\\
    \includegraphics[trim=1mm 4mm 1mm 3mm, clip, width=\textwidth]{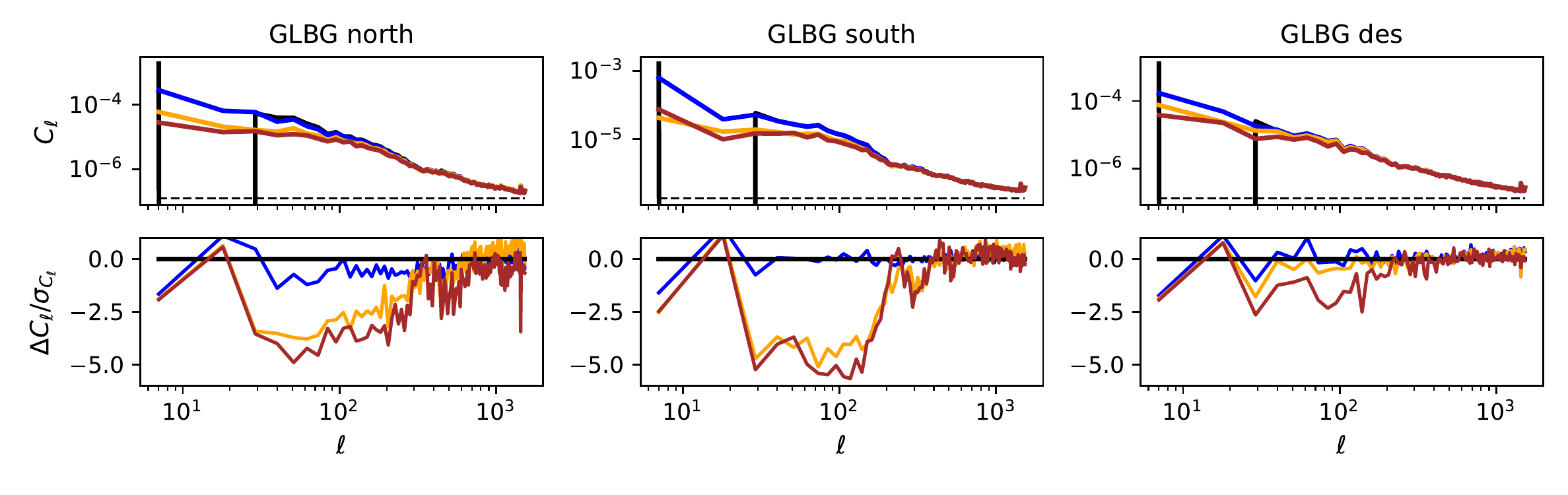}
    \caption{Angular power spectra of the uncorrected and corrected galaxy overdensity maps obtained with the different cleaning methods compared. The bottom panels show the fractional changes, i.e the
    corrected spectra minus the uncorrected one, divided by Gaussian uncertainties defined in \secref{sec:maps_and_power_spectra}. While the changes on large scales are the most pronounced, the cleaning results in a reduction of power on almost all scales, and the removal of sharp contamination features (\eg $\ell \sim 300$).}
    \label{fig:power_spectra}
\end{figure*}

\renewcommand{\thefigure}{\arabic{figure} (Continued)}
\addtocounter{figure}{-1}

\begin{figure*}[!ht]
    \centering
    \includegraphics[trim=1mm 4mm 1mm 3mm, clip, width=\textwidth]{figures/power_spectra_qso.pdf}\\
    \includegraphics[trim=1mm 4mm 1mm 3mm, clip, width=\textwidth]{figures/power_spectra_GLBG.pdf}
    \caption{}
\end{figure*}

\subsection{Original, corrected maps, and angular power spectra}

\figref{fig:maps_orig} shows the original, uncorrected galaxy number counts.
It can be seen that: 
1) Except for LRGs, high levels of spurious spatial variations are present.
2) The depth and spatial variations of some of the samples can be significantly different between the three regions.
This follows previous findings \citep{chaussidon2021angular, Kitanidis_2020, Zarrouk_2021}: 
LRGs have little contamination;
ELGs are much more highly contaminated due to selection near the depth limit;
QSOs also exhibit high contamination, which follows the stellar density since stars and quasars are easily confused.

\figref{fig:maps_deep_corr} shows the maps corrected with the deep sets method. 
Very little contamination-like patterns can be observed.
They all show comparable levels of quasi-uniform and isotropic fluctuations.
\figref{fig:maps_nn} shows the correlations extracted by the neural network technique, and \figref{fig:maps_deep} the additional correlations found by the deep sets technique.
This allows us to examine more closely what patterns in the number counts were successfully explained by the regression as a function of the potential contaminants.

In order to gain further insight, we examine the angular power spectra (band-powers), since they are a summary statistic from which cosmological information is typically extracted, and harmonic modes (the spherical coordinates equivalent of Fourier modes) help separate effects on different physical scales for signals on the sphere.

\figref{fig:power_spectra} shows the angular power spectra of the original and corrected galaxy catalogs (number counts converted to over-density maps).
The sub-panels show the power subtracted by each method, normalized by the Gaussian (diagonal) error.
We see that:
\begin{itemize}
    \item Overall, power is removed by all three methods. This is consistent with the expectation that the intrinsic cosmological signal is isotropic and close to uniform, with little power on large scales, and that any contaminant will add large amounts of excess power.
    \item The more complex and flexible the model, the more power is identified and removed, as one would expect in the presence of significant contamination. This is aligned with the results of Table~\ref{tab:results}.
\end{itemize}

\subsection{Chance correlations}

Chance correlations are random alignments between the cosmological signal and the possible contaminants.
They are a potential issue with all methods discussed here.
For example, the power spectrum (additive) bias that results from fitting the linear model to $n$ contaminant maps is $b_\ell \sim - n /( \fsky^2 (2\ell + 1))$, as derived in \cite{Elsner_2015}
This bias increases with the numbers of degrees of freedom of the model and the inputs, and is therefore expected to be larger for the non-linear and deep sets models.
However, the chance correlations bias resulting from fitting each model mostly depends on the inputs (contaminant maps for the linear and neural network model, and stacks for the deep sets) and on the statistical properties of the sample under consideration, mostly its power spectrum.
Given that the power spectra of the galaxy samples are similar, and that the samples themselves exhibit some correlation due to their overlap in redshift, we can in fact obtain a rough limit for the chance correlations bias in the LRG power spectra.
Indeed, it is the cleanest data, by construction, and thus the case where the bias is likely to be most visible.
In all other cases, the subtracted power is so much larger that it is unlikely to be due to chance correlations, but rather to contamination.
However, simulations would be needed to confirm this upper limit on the bias.

\renewcommand{\thefigure}{\arabic{figure}}

\begin{figure*}[!ht]
    \centering
    \includegraphics[width=0.47\textwidth]{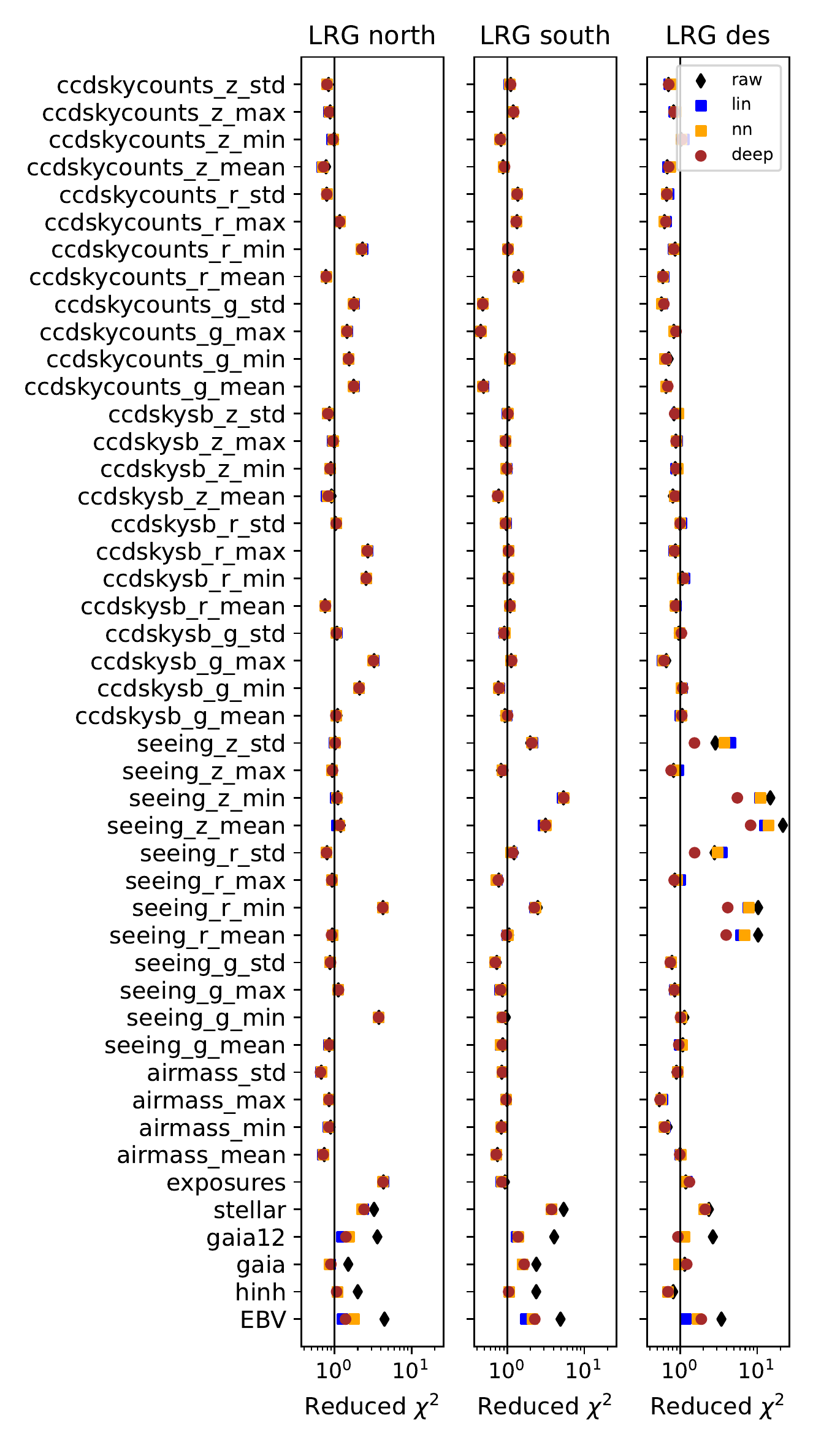}
    \includegraphics[width=0.47\textwidth]{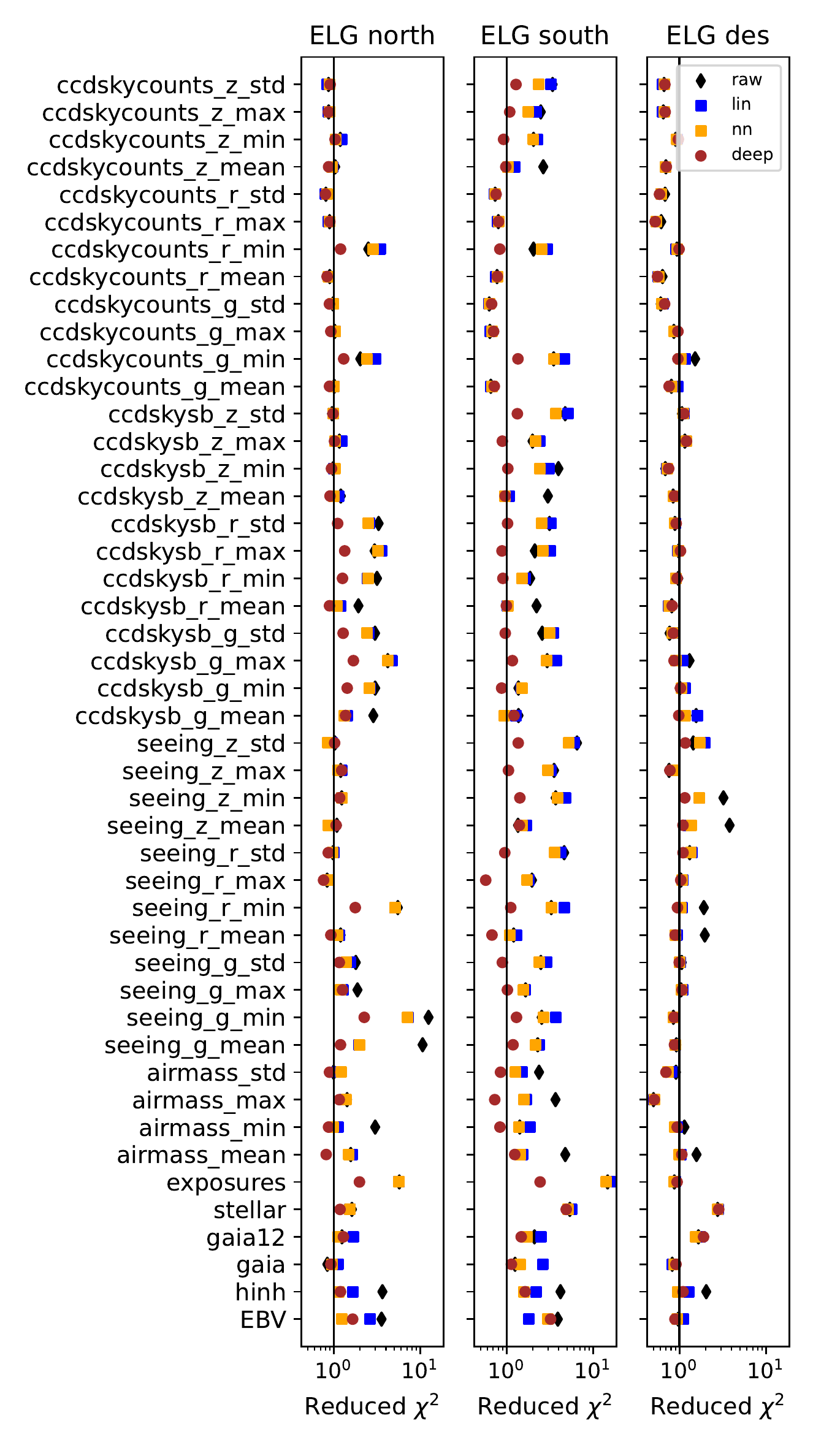}\\
    \caption{Residual contamination as measured in cross-power spectra between cleaned galaxy overdensity  maps and contaminants maps. The reduced chi-squared is calculated by comparing the cross-spectra (up to $\lmax=1024$) with a zero power spectrum as a hypothesis, using the procedure detailed in \secref{sec:maps_and_power_spectra}.}
    \label{fig:syschi2s}
\end{figure*}

\renewcommand{\thefigure}{\arabic{figure} (Continued)}
\addtocounter{figure}{-1}

\begin{figure*}[!ht]
    \centering
    \includegraphics[width=0.47\textwidth]{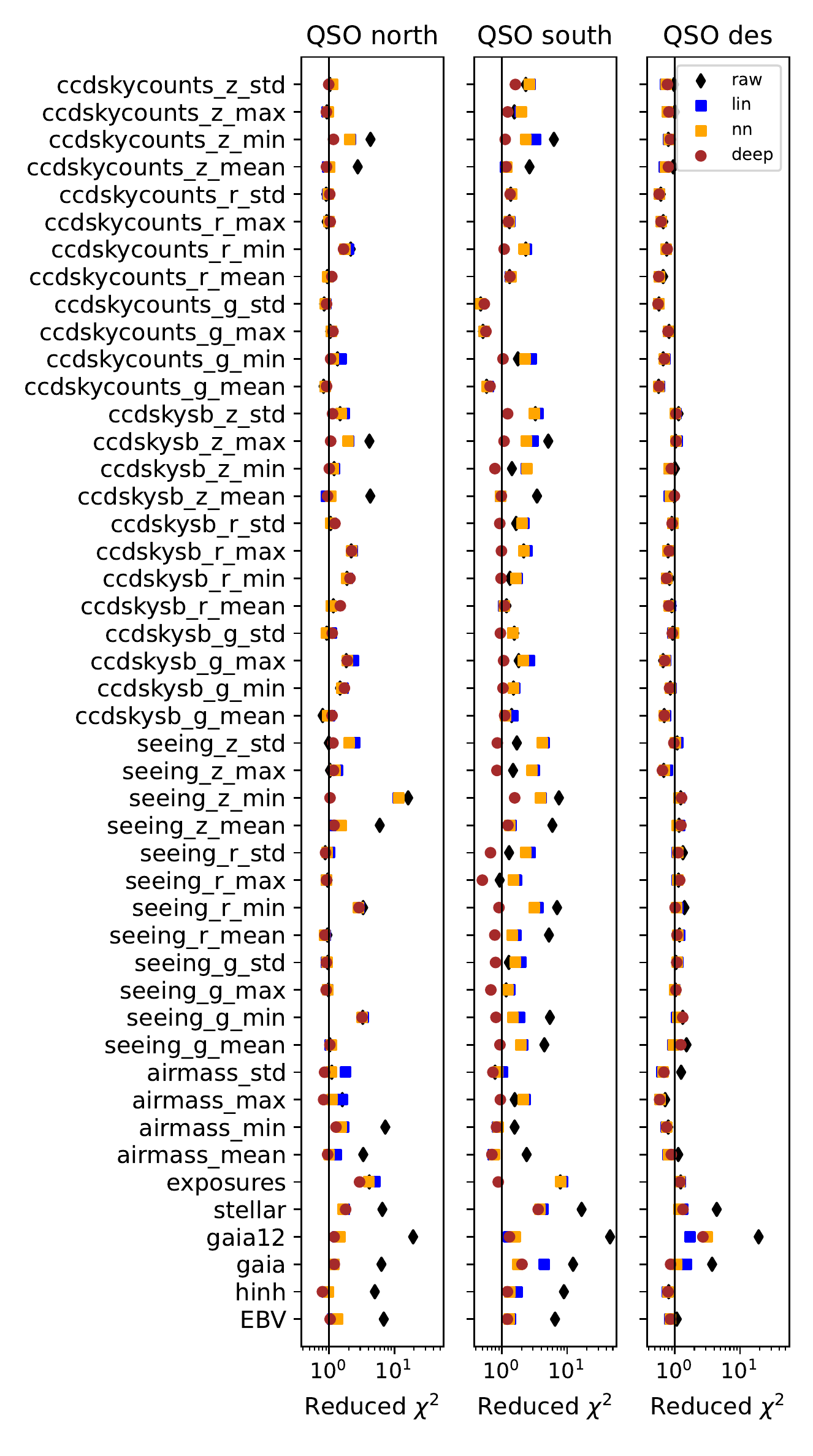}
    \includegraphics[width=0.47\textwidth]{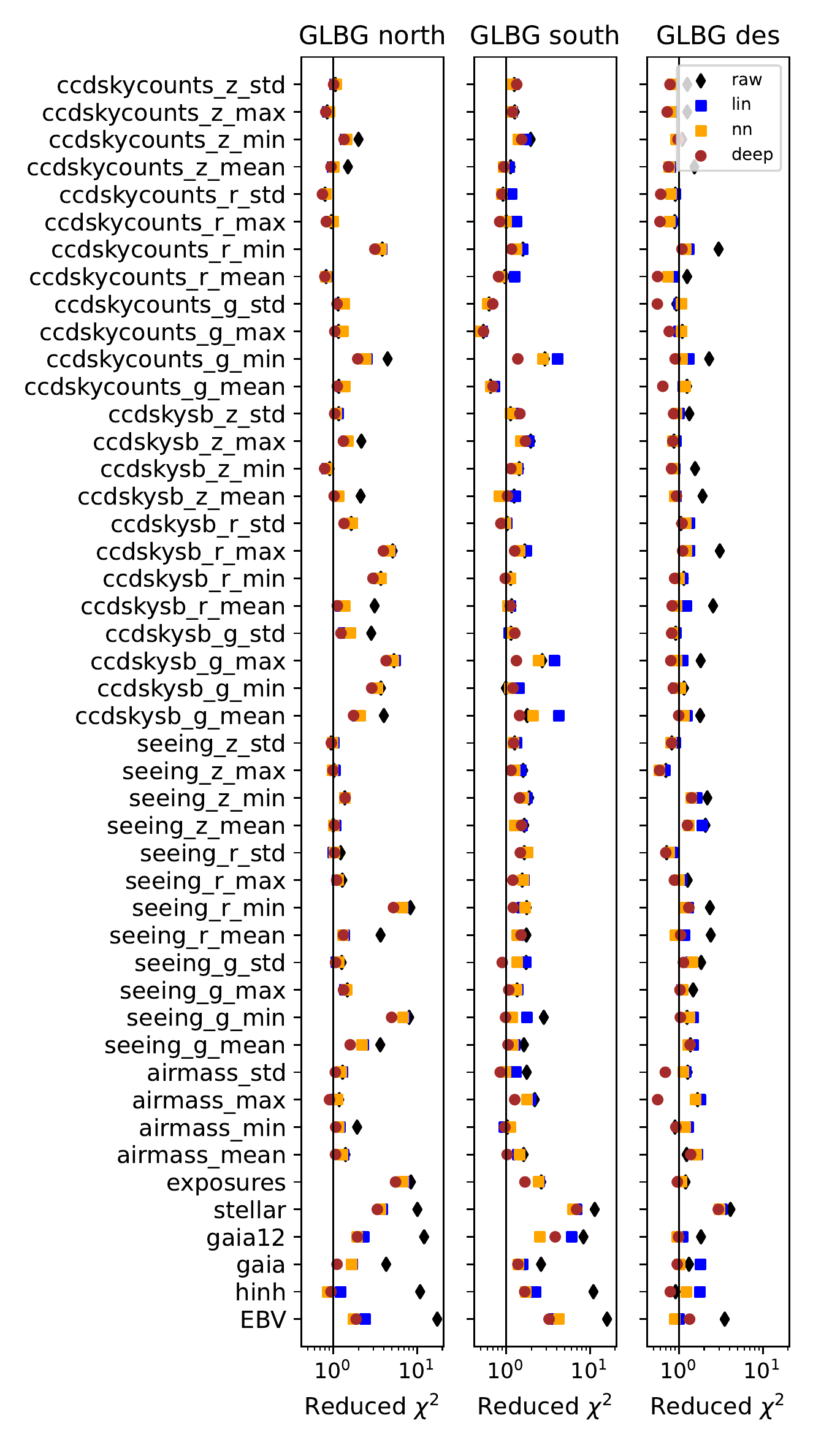}\\
    \caption{}
\end{figure*}

\renewcommand{\thefigure}{\arabic{figure}}

\subsection{Null tests cross-power spectra} 

In order to gain more intuition about the sources of the power removed by each technique, we calculate cross-power spectra between contaminant maps and number density maps.
However, the original contaminants maps used in the neural network technique are, by construction lossy compressions of the exposure stacks used in the deep sets technique. 
Cross-correlations with those maps (e.g., mean seeing) will reveal if the deep sets fits are indeed able to remove more contamination from those sources, as expected.
Yet, it would be interesting to measure by how much it beats the neural network for information that the latter has not access to.
For this purpose, we build an other set of contaminant maps, with summary statistics (the minimum, maximum, and standard deviation) of the exposure stacks.
We also add a map of the number of exposures (not feed directly to any of the methods).
While they could in theory be fed into the linear and neural network methods, it is unclear if those are good summary statistics, and the point of the deep sets is to directly capture all the information from the stacks and circumvent the need to pick summary statistics.

We compute cross-angular (band-)power spectra  $\{\mathcal{C}_{b}\}_{b=1,\cdots, B}$ between the galaxy overdensity maps and the new extended set of contaminants maps. We then calculate covariance matrices $\{C_{bb^\prime}\}$ with the procedure detailed in \secref{sec:maps_and_power_spectra}.
We compute chi-squared values $\chi^2 = \sum_{b=1}^B \sum_{b^\prime=1}^B C^{-1}_{bb^\prime} \mathcal{C}_{b} \mathcal{C}_{b^\prime}$, which correspond to the hypothesis that the power spectra are consistent with zero, \ie no detectable contamination signal. 
We reduce the chi-squared by dividing them with the number of harmonic multipole bands.

\figref{fig:syschi2s} shows the resulting reduced chi-squared values.
The value of one, corresponding to the hypothesis (zero power spectrum) for the limit of no contamination, is indicated with vertical lines.
Even though this calculation relies on the assumption that the fields are Gaussian, and also on fiducial theoretical power spectra, this figure can be used to obtain a rough quantification of which sources of contamination are most significant in each sample, and how well each method performs.

Across the board, the deep sets method removes more correlations than the other methods.
We should consider the contaminants that were accessible to the neural network and linear models, \ie the mean maps. 
In that case, the performance of the neural network and the deep sets is generally comparable. However there are a few cases where the deep sets performs better. 

When it comes to the contaminant maps that were not given to the linear or neural network models (standard deviation, min, max), the deep sets substantially reduces the measured residual correlation. 
This is particular dramatic for some of the samples, which gives an indication of the possible sources of contamination.
Yet, in cases where significant correlations are present in the raw data, the improvements provided by the deep sets model are significant, but do not systematically bring the reduced chi-squared close to unity.
 

There is significant residual contamination in a range of cases, which the deep sets should have been able to extract, in theory.
However, this could be because the models were only trained on a fraction of the data/sky.
Training the model on the full data set may resolve this.
But it could result in increased overfitting and removal of cosmological power. 
Thus this may require more in-depth study of chance correlations and overfitting.
In order to reduce this bias, we have taken the approach to fit the models on a fraction of the data, splitting into training and testing sets. 
Other recent works have shown that this is unbiased in the cases studied \citep{Rezaie_2020, Zarrouk_2021, Kitanidis_2020}.
This approach warranted a more robust fitting and optimization of our hyperparameters, and a fairer comparison between the models.
However, it also means correlations with the potential contaminants are not necessarily captured at their full extent, which may then leave some residual systematic biases.
Another way to improve the robustness of corrections techniques and to prevent over fitting and chance correlations is to add terms to the loss function. For example, one can account for the expected covariances (between pixels) due to cosmological fluctuations \cite{Wagoner_2021}. We defer those points to future work.

\section{Discussion and conclusion}\label{sec:conclusion}

Modern galaxy surveys are plagued by observational systematics, which bias cosmological analyses if not mitigated.
One major source of such systematic biases is the spurious modulation of galaxy properties (\eg number counts) due to varying observing conditions. 
While this is now routinely treated with linear and non-linear models based on contaminant template maps, this approach neglects the multi-epoch nature of those surveys.
This work is motivated by the fact that this has not been studied for wide field surveys.
Extracting this information could lead to cleaner galaxy catalogs and more accurate cosmological analyses.

We presented a new method that can tap into the information in image stacks.
We restricted our attention to the metadata of the exposures (seeing, airmass, etc) in the exposure stacks to predict the galaxy number counts.
However, learning the contamination from the raw images themselves should be possible and is left for future work.  

The new method relies on deep sets architectures, modified to better suit pixelized galaxy number counts, variable-length image stacks, and additional contaminant sky maps.  
We applied this method to four types of galaxies extracted from catalogs and images from the DESI Legacy Surveys.
Those exhibit different levels of contamination. 
The new model is capable of significantly reducing this contamination, to a greater extent than conventional linear or non-linear models.
Since we do not go all the way to cosmological parameters, we have not performed more exhaustive null tests with other metrics.

An other promising avenue for optimizing the analysis of galaxy surveys is to define sky cuts achieving an optimal balance between loss of information and accuracy of the systematics mitigation. 
This could be done with the results presented here.

The DESI Legacy Surveys have a few (to tens) of single-epoch images per band contributing to each sky location, on average.
This number will dramatically increase for LSST. 
This may improve survey uniformity, since more images are contributing to each detection, and average survey properties would be better summary statistics. 
But this may also increase the likelihood of extreme outliers contributing.
The result may therefore go either way in terms of effective contamination on the sky. 
Thus, a dedicated study of deep sets models would be needed for LSST, to clarify if methods working with image stacks directly are indeed necessary.

DESI will soon provide follow-up spectroscopy for a lot of the objects analysed in this work. 
This will lead to secure measurements of galaxy type, redshift, and could even fix erroneous photometry.
Yet, it is known that the spurious correlations in the target photometry can propagate into the spectroscopic catalogs (\eg \citealt{Ross_2011, Rezaie_2021}), and therefore the techniques explored here are relevant to both photometric and spectroscopic analyses.

Finally, we note that the injection of additional sources (so called Synthetic Source Injections, SSI) in images (single-epochs or coadds) is a powerful technique to simulate contamination, and to validate systematics mitigation techniques in cosmological analyses.
It has been key in recent cosmological analyses and will play an increased role in the LSST era \citep{Everett_2022, Huang_2017}, complementary to the techniques shown in this paper.

\newpage
\textbf{Acknowledgements.} 

EE and BL thank the members of the cosmology and quasars groups at Imperial College for useful discussions and feedback during the course of this project.

BL is supported by the Royal Society through a University Research Fellowship.

The Legacy Surveys consist of three individual and complementary projects: the Dark Energy Camera Legacy Survey (DECaLS; Proposal ID 2014B-0404; PIs: David Schlegel and Arjun Dey), the Beijing-Arizona Sky Survey (BASS; NOAO Prop. ID 2015A-0801; PIs: Zhou Xu and Xiaohui Fan), and the Mayall z-band Legacy Survey (MzLS; Prop. ID 2016A-0453; PI: Arjun Dey). DECaLS, BASS and MzLS together include data obtained, respectively, at the Blanco telescope, Cerro Tololo Inter-American Observatory, NSF’s NOIRLab; the Bok telescope, Steward Observatory, University of Arizona; and the Mayall telescope, Kitt Peak National Observatory, NOIRLab. The Legacy Surveys project is honored to be permitted to conduct astronomical research on Iolkam Du’ag (Kitt Peak), a mountain with particular significance to the Tohono O’odham Nation.

NOIRLab is operated by the Association of Universities for Research in Astronomy (AURA) under a cooperative agreement with the National Science Foundation.

This project used data obtained with the Dark Energy Camera (DECam), which was constructed by the Dark Energy Survey (DES) collaboration. Funding for the DES Projects has been provided by the U.S. Department of Energy, the U.S. National Science Foundation, the Ministry of Science and Education of Spain, the Science and Technology Facilities Council of the United Kingdom, the Higher Education Funding Council for England, the National Center for Supercomputing Applications at the University of Illinois at Urbana-Champaign, the Kavli Institute of Cosmological Physics at the University of Chicago, Center for Cosmology and Astro-Particle Physics at the Ohio State University, the Mitchell Institute for Fundamental Physics and Astronomy at Texas A\&M University, Financiadora de Estudos e Projetos, Fundacao Carlos Chagas Filho de Amparo, Financiadora de Estudos e Projetos, Fundacao Carlos Chagas Filho de Amparo a Pesquisa do Estado do Rio de Janeiro, Conselho Nacional de Desenvolvimento Cientifico e Tecnologico and the Ministerio da Ciencia, Tecnologia e Inovacao, the Deutsche Forschungsgemeinschaft and the Collaborating Institutions in the Dark Energy Survey. The Collaborating Institutions are Argonne National Laboratory, the University of California at Santa Cruz, the University of Cambridge, Centro de Investigaciones Energeticas, Medioambientales y Tecnologicas-Madrid, the University of Chicago, University College London, the DES-Brazil Consortium, the University of Edinburgh, the Eidgenossische Technische Hochschule (ETH) Zurich, Fermi National Accelerator Laboratory, the University of Illinois at Urbana-Champaign, the Institut de Ciencies de l’Espai (IEEC/CSIC), the Institut de Fisica d’Altes Energies, Lawrence Berkeley National Laboratory, the Ludwig Maximilians Universitat Munchen and the associated Excellence Cluster Universe, the University of Michigan, NSF’s NOIRLab, the University of Nottingham, the Ohio State University, the University of Pennsylvania, the University of Portsmouth, SLAC National Accelerator Laboratory, Stanford University, the University of Sussex, and Texas A\&M University.

BASS is a key project of the Telescope Access Program (TAP), which has been funded by the National Astronomical Observatories of China, the Chinese Academy of Sciences (the Strategic Priority Research Program “The Emergence of Cosmological Structures” Grant XDB09000000), and the Special Fund for Astronomy from the Ministry of Finance. The BASS is also supported by the External Cooperation Program of Chinese Academy of Sciences (Grant 114A11KYSB20160057), and Chinese National Natural Science Foundation (Grant 11433005).

The Legacy Survey team makes use of data products from the Near-Earth Object Wide-field Infrared Survey Explorer (NEOWISE), which is a project of the Jet Propulsion Laboratory/California Institute of Technology. NEOWISE is funded by the National Aeronautics and Space Administration.

The Legacy Surveys imaging of the DESI footprint is supported by the Director, Office of Science, Office of High Energy Physics of the U.S. Department of Energy under Contract No. DE-AC02-05CH1123, by the National Energy Research Scientific Computing Center, a DOE Office of Science User Facility under the same contract; and by the U.S. National Science Foundation, Division of Astronomical Sciences under Contract No. AST-0950945 to NOAO.

\bibliography{bib}{}
\bibliographystyle{aasjournal}


\end{document}